\documentstyle[epsf]{article}
\def\m@thcombine#1#2{%
  \setbox0=\hbox{$#1$}
  \setbox1=\hbox{$#2$} 
  \ifdim\wd0>\wd1
    \setbox0=\hbox to\wd1{\hss\box0\hss}
  \else
    \setbox1=\hbox to\wd0{\hss\box1\hss}
  \fi
  \mathop{\vcenter{
    \offinterlineskip\box0\box1}}}
\def\lesim{\m@thcombine<\sim}
\def\gesim{\m@thcombine>\sim}
\def\lessgtr{\m@thcombine<>}
\def\gtrless{\m@thcombine><}
\makeatletter
\def\m@thcombine#1#2{%
  \setbox0=\hbox{$#1$}
  \setbox1=\hbox{$#2$}
  \ifdim\wd0>\wd1
    \setbox0=\hbox to\wd1{\hss\box0\hss}
  \else
    \setbox1=\hbox to\wd0{\hss\box1\hss}
  \fi
  \mathop{\vcenter{
    \offinterlineskip\box0\box1}}}
\def\lesim{\m@thcombine<\sim}
\def\gesim{\m@thcombine>\sim}
\def\lessgtr{\m@thcombine<>}
\def\gtrless{\m@thcombine><}
\makeatother

\newcommand{\ket}[1]{\left| #1 \right\rangle}
\newcommand{\omegaI}{\omega_I}

\newcommand{\beq}{\begin{equation}}
\newcommand{\beqa}{\begin{eqnarray}}
\newcommand{\eeq}{\end{equation}}
\newcommand{\eeqa}{\end{eqnarray}}

\begin{document}

\LARGE

\begin{center}
{\bf  Shell Effects on Rotational Damping in Superdeformed Nuclei
} 
\end{center}

\large
\vspace{5mm}

\begin{center}
K. Yoshida and  M. Matsuo$^{a}$,
\end{center}

\vspace{5mm}

\large

\begin{center}
{\it
Research Center for Nuclear Physics, Osaka University,
Ibaraki, Osaka, Japan\\
${}^{a}$Yukawa Institute for Theoretical Physics, Kyoto University,
Kyoto 606-01, Japan \\
}
\end{center}

\vspace{5mm}

\normalsize

\begin{abstract}
Damping of collective rotational motion in $A\sim 190$  and $A\sim
150$ superdeformed nuclei is studied by means of the cranked shell
model with 
two-body residual force. Numerical calculations predict 
that in a
typical $A\sim 190$ superdeformed nucleus, $^{192}$Hg, the
rotational 
damping width is significantly small, $\Gamma_{\rm rot}\sim
30$ keV, and that the number of superdeformed bands in the 
off-yrast region amounts up to 150 at a given rotational frequency.
These features are quite different from the prediction for $A\sim 150$
superdeformed nuclei and rare-earth normally deformed nuclei. It is
shown that the single-particle alignments of the cranked Nilsson
orbits have strong shell oscillation. It  affects
significantly the properties of rotational damping in superdeformed
$^{192}$Hg.
\end{abstract}

{\it keywords} : Rotational damping, Superdeformed nuclei, Shell effect,
Angular momentum alignment

\section{Introduction}\label{sec:intro}
Recent progress in $\gamma$-ray spectroscopy has enabled us to 
study off-yrast levels and associated E2 transitions in 
rapidly rotating nuclei.
The observed rotational bands lie near the yrast line and they are
well described as
quasi-particle or particle-hole excitations in a rotating mean
field.
Excited states which lie at higher intrinsic excitation energy 
above the yrast line are hardly observed by means of discrete
$\gamma$-ray spectroscopy not only because individual levels are
populated very weakly but also because they 
do not necessarily form the rotational
band structure due to damping of collective rotational motion
\cite{Leander82,Lauritzen86}.
The damping of collective rotational motion takes place in the energy
region where density of
many-particle many-hole states in the rotating mean
field is high and the residual
interaction causes configuration mixing
among them. If the rotational damping sets in, E2 decay strength from
an excited 
state spreads out over many final states.  The 
number of rotational bands existing in a given nucleus is thus
limited. 
The rotational damping  has recently been studied experimentally
through the 
statistical analyses of the
quasi-continuum $\gamma$-ray spectra emitted from high spin off-yrast
states. The fluctuation analysis
method(FAM) \cite{Herskind92}
which extracts effective 
number of E2 decay paths revealed that in $A\sim 170$
rare-earth normally deformed nuclei there are only about 30 
rotational band structures at a given rotational frequency
and the rotational damping sets in at around
0.8 MeV above yrast line.
 The occurence of rotational damping is reproduced 
by a microscopic realistic calculation, which takes into
account the configuration mixing in the rotating mean field by means
of the shell model diagonalization method \cite{Aberg92,Matsuo97}.

It has been argued theoretically that  a 
strong shell effect is expected on the rotational damping in
superdeformed nuclei \cite{Aberg92,Yoshida97}
since the presence of the shell gap in the
single particle spectrum 
decreases dramatically the level
density of excited many-particle many-hole configurations.
This shell effect
influences the intrinsic excitation energy where the rotational
damping sets in. In a previous paper \cite{Yoshida97} we
performed a systematic analysis for
superdeformed nuclei in $A\sim 150$ region by means of the cranked
shell model approach and predicted that the onset 
energy of damping and number of rotational bands are higher/larger than
in $A\sim 170$ normally deformed nuclei, {\it i.e.},
$E_{\rm onset}=2\sim 3$ MeV and $N_{\rm band}=30\sim 100$,
and  they vary significantly as functions 
of neutron and proton numbers  due to the change in the shell
gap. 
The continuum E2 $\gamma$-ray spectra associated with superdeformed
states are observed experimentally in
$^{152}$Dy \cite{Schiffer91}, $^{143}$Eu \cite{Leoni95} and
$^{192}$Hg \cite{Lauritsen92}. So far the
FAM analysis has been applied to $^{143}$Eu \cite{Leoni95} for which
the theoretical 
prediction gives
a qualitatively consistent account of the number of rotational bands
$N_{\rm band}$.  

In this paper we will reveal
that in addition to the  level density  effect there exists another kind of
shell effect which affects significantly the rotational damping 
properties,
especially the damping width, which is another fundamental quantity
characterizing the rotational damping besides $N_{\rm band}$ and
$E_{\rm onset}$. To demonstrate this, we investigate in detail
off-yrast superdeformed states 
and associated rotational damping properties in a typical
superdeformed nucleus $^{192}$Hg in  the  $A\sim 190$ region, and
make comparative study with  $A\sim 150$ superdeformed nuclei and
a typical normally deformed rare-earth nucleus $^{168}$Yb.  The shell
effect discussed here is originated from a 
shell structure associated with the angular momentum alignment of the
single-particle orbits in the  cranked  mean field. 
As was discussed previously \cite{Nishinomiya} highly aligned 
single-particle orbits affect
the dispersion in rotational frequency of cranked many-particle
many-hole states, which is a key quantity governing the
rotational damping width. 
In this paper, we will show that
because of the new shell 
effect the rotational damping width in superdeformed 
$^{192}$Hg becomes remarkably
small. As a consequence, the mechanism of onset of
rotational damping in superdeformed $^{192}$Hg becomes slightly
different from that in $A\sim 150$ superdeformed and $A\sim 170$
normally deformed nuclei.

To this end, we perform a fully microscopic description of the
superdeformed 
states in the high spin off-yrast region of $^{192}$Hg by means of the
shell model method based on the cranked mean field \cite{Matsuo97}.
Since the formulation is essentially the same as the
previous paper \cite{Yoshida97}, it is described only briefly in
Sec.\ref{sec:form}. In Sec.\ref{sec:results} we show results obtained by our numerical
calculation and characterize the onset
of rotational damping and rotational damping width for 
$^{192}$Hg.  In Sec.\ref{sec:discuss}
we discuss the results in order to understand mechanism of striking
features 
in rotational damping in $^{192}$Hg in terms of the shell effect
associated with the single particle alignment structure.
In Sec.\ref{sec:conclude} we will make a summary on the shell effects
on rotational damping in superdeformed nuclei clarified in the
previous and present  investigations. 

\section{Formulation}\label{sec:form}
To describe excited superdeformed states in the off-yrast region in
$^{192}$Hg, we first construct unperturbed basis wave functions which
are defined as 
multi-particle multi-hole
configurations generated upon diabatic single particle routhian
orbits \cite{Bengtsson89} of cranked  
Nilsson mean-field Hamiltonian $h^{\omega}=h_{\rm Nilsson}-\omega j_x$. 
Here deformation parameters of the Nilsson potential are determined by
means of Strutinsky minimization scheme \cite{Anderson76} with the
pairing correlation  energy  included.
We consider quadrupole and
hexadecupole deformation parameters ($\epsilon_2,\epsilon_4$).
We determine  equilibrium deformation at zero 
rotational frequency.
The  shell correction energy for the pairing correlation is included
in a standard way \cite{Brack72}, $E^{\rm pair}=-(P-\tilde{P})$,
where $P$ is pairing correlation energy gain obtained by the BCS
approximation and $\tilde{P}$ is the smoothed part
obtained with  
equidistant single particle levels.
Single particle levels within $\pm 6$ MeV from the Fermi surface 
are considered. The pairing coupling constant $G$ is determined so
that the 
pairing gap for the equidistant model gives 
$12/A^{-1/2}$ MeV.
The total energy, including the liquid
drop energy, shell correction energy and $E^{\rm pair}$, evaluated
at zero rotational frequency is minimized with respect to
$\epsilon_2$ and $\epsilon_4$.
Thus we obtained the equilibrium
deformation parameters
$(\epsilon_2,\epsilon_4)=(0.451,0.0115)$ for superdeformed state in
$^{192}$Hg. The same deformation parameters are used to construct the
basis states at high spins
because the 
deformation is considered to change only slightly with increasing spin.

Total energy of a basis $n$p-$n$h configuration $\mu$ is expressed as 
\begin{eqnarray}
\label{basis-e}
E_{\mu}(I)&=&
\sum_{
\small
\begin{array}{c}
{\rm occupied}\;n \\
{\rm in}\;\mu
\end{array}
}
 e_n(\omegaI) 
+\omegaI I+\frac{1}{2 {\cal
J}_{\mu}^{\rm micro}}(I-J_{x\mu}(\omegaI))^2-E_{\rm smooth}(I)+E^{\rm
RLD}(I),\\
J_{x\mu}&=&\sum_{
\small
\begin{array}{c}
{\rm occupied}\;n \\
{\rm in}\;\mu
\end{array}
}
i_n(\omegaI)
\end{eqnarray}
where  $e_n(\omegaI)$  is single particle routhian energy for the
cranked Nilsson 
Hamiltonian.
$J_{x\mu}(\omega_I)$ and ${\cal J}_{\mu}^{\rm micro}$ are
expectation value of the angular momentum 
operator $\hat{J_x}$ and 
dynamic moment of inertia, respectively,
for the many-body configuration $\ket{\mu(I)}$. $i_n(\omegaI)$ is the
$\hat{j}_x$ expectation value for the cranked Nilsson single particle
orbits. Here reference 
rotational frequency 
$\omega_I$ is 
determined so that $J_x(\omega_I)=I$ is satisfied for  the  yrast
configuration. 
$E^{\rm smooth}$ and $E^{\rm RLD}$ are the smooth part 
and the rotating liquid drop energy \cite{Myers-Swiatecki67},
respectively.

As the residual two-body interaction we adopt the delta interaction,
$v(1,2) = -v_{\tau}\delta ({\vec{x}}_1 - {\vec{x}}_2)$,
$v_{\tau}= 340 {\rm MeV fm}^3$
for like nucleons and $v_{\tau}= 500 {\rm MeV fm}^3$ for the
neutron-proton interaction \cite{Bush92}.
A shell model Hamiltonian matrix for the $n$p-$n$h basis states
$\{\ket{\mu(I)}\}$ is constructed by including the residual two-body
interaction and 
diagonalized numerically for each spin and parity.
The obtained energy eigenstates  are  linear combinations of
unperturbed basis,
\begin{equation}
\ket{\alpha(I)}=\sum_\mu X_{\mu}^{\alpha}(I)\ket{\mu(I)}.
\end{equation}
The E2 transition strength between an initial state $\alpha$ at $I$
and a final state $\beta$ at $I-2$ is given by 
\begin{equation}
M^2_{\alpha I\beta I\! -\! 2}\\
=\frac{15}{128\pi}Q_0^2 \mid \sum_\mu X^{\beta *}_\mu (I-2)
X^\alpha_\mu (I)\mid^2,
\end{equation}
assuming that collective E2 transition takes place between
the basis states having the same microscopic configuration and
the static quadrupole moment $Q_0$ is constant for all
configurations.
Then 
\begin{equation}
w_{\alpha I\rightarrow\beta I-2}=\mid \sum_\mu X^{\beta *}_\mu (I-2)
X^\alpha_\mu (I)\mid^2
\end{equation}
represents normalized transition strength
satisfying  
$\sum_{\beta}w_{\alpha I\rightarrow\beta I-2}=1$.

The single particle levels located within
4.0 MeV above/below the Fermi surface are taken into account as active 
orbits in constructing the many-body basis states, and up to
five-particle five 
-hole configurations are included. The energetically lowest 2000 basis
states (for 
each $I^{\pi}$)  
are chosen to diagonalize the Hamiltonian matrix. The model space of
the diagonalization is not large enough to describe static pairing
correlation which  may be  important for near-yrast states at low spins. It 
is to be noted, however, that the pairing correlation is weakened by
the Coriolis anti-pairing effect at very high spin and also by the
blocking effect for the excited superdeformed states in the off-yrast
region. 

In the following section we also show results for superdeformed states
in $^{143}$Eu and $^{152}$Dy, selected as representatives in  the  $A\sim
150$ region, and also normally deformed high spin states in a
typical rare-earth nucleus $^{168}$Yb. The calculational procedure for
those nuclei were already described in the previous
publications \cite{Yoshida97,Matsuo97}. The numerical  
results shown in this paper are obtained by using the same delta
interaction  
as in $^{192}$Hg.

It is known that the superdeformed states usually coexist with
weakly deformed states in the same nucleus at the same spin
and parity.
They do not mix, however, very strongly at high spin regions where
there is a potential barrier 
between the two kinds of states and the level density of weakly
deformed states is relatively small.
To make comparison with experimental data including low spin
region where the mixing may be effective, it is necessary to take into
account the barrier penetration. The excited superdeformed states may
decay not only via 
collective E2 but also via statistical E1 transition which becomes
possible through the mixing of weakly deformed
states \cite{Vigezzi90,Shimizu92}. 
Since our aim in this paper is to discuss mechanism of rotational
damping we leave this effect for future investigation.

\section{Results}
\label{sec:results}
\subsection{Onset of rotational damping}
In this subsection, we discuss 
onset property of rotational damping by analyzing onset energy of
rotational damping and 
number of superdeformed rotational bands.
\begin{figure}
\epsfxsize=150mm\epsffile{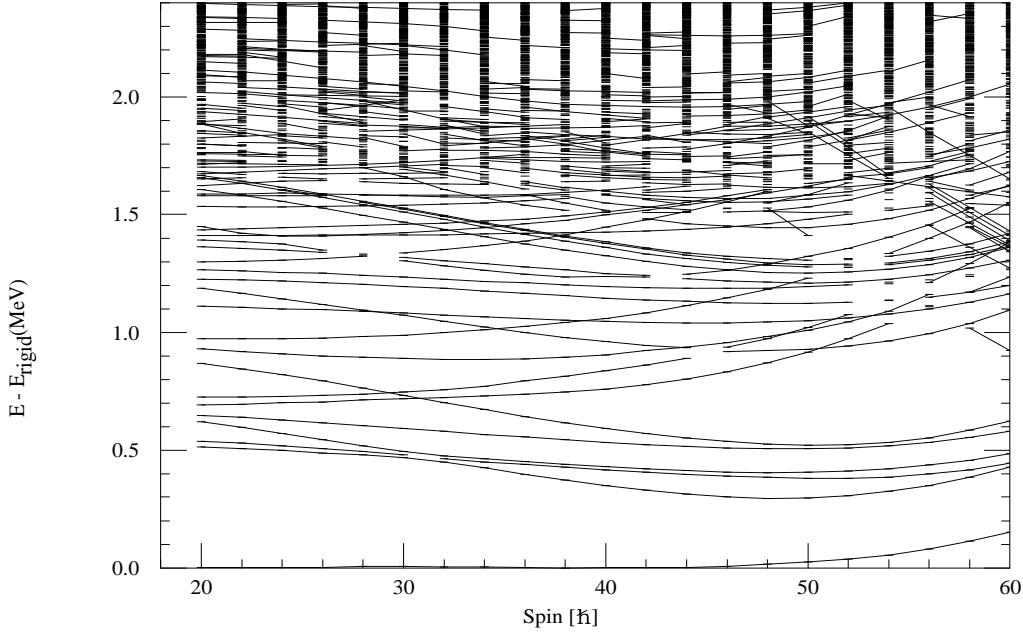}
\caption
{\label{flow}
The calculated energy levels plotted
with little horizontal bars for $(\alpha,\pi)=(0,+)$ states in
superdeformed 
$^{192}$Hg. A rigid-rotor 
rotational energy 
$E_{\rm rigid} = I(I+1)/234$ (MeV) is
subtracted.  
Solid lines connecting the energy levels represent strong E2
transitions with normalized strength larger than $ \protect
\sqrt{1/2}=0.707$. 
}
\end{figure}
In Fig.\ref{flow} we display obtained superdeformed levels for
parity $\pi=+$, signature $\alpha=0$ in $^{192}$Hg.
Energy levels are shown by short horizontal bars. 
Strong E2
transitions satisfying $w_{\alpha I\rightarrow \beta I-2}\ge
0.707=\frac{1}{\sqrt{2}}$ are plotted with solid lines, which are 
discrete transitions forming the rotational band structure.
The rigid-body rotor energy $E_{\rm rigid}=I(I+1)/234$ MeV is
subtracted. The vertical
axis indicates approximately the intrinsic excitation energy measured
from the yrast line.

\begin{figure}
\centerline{
\epsfxsize=70mm\epsffile{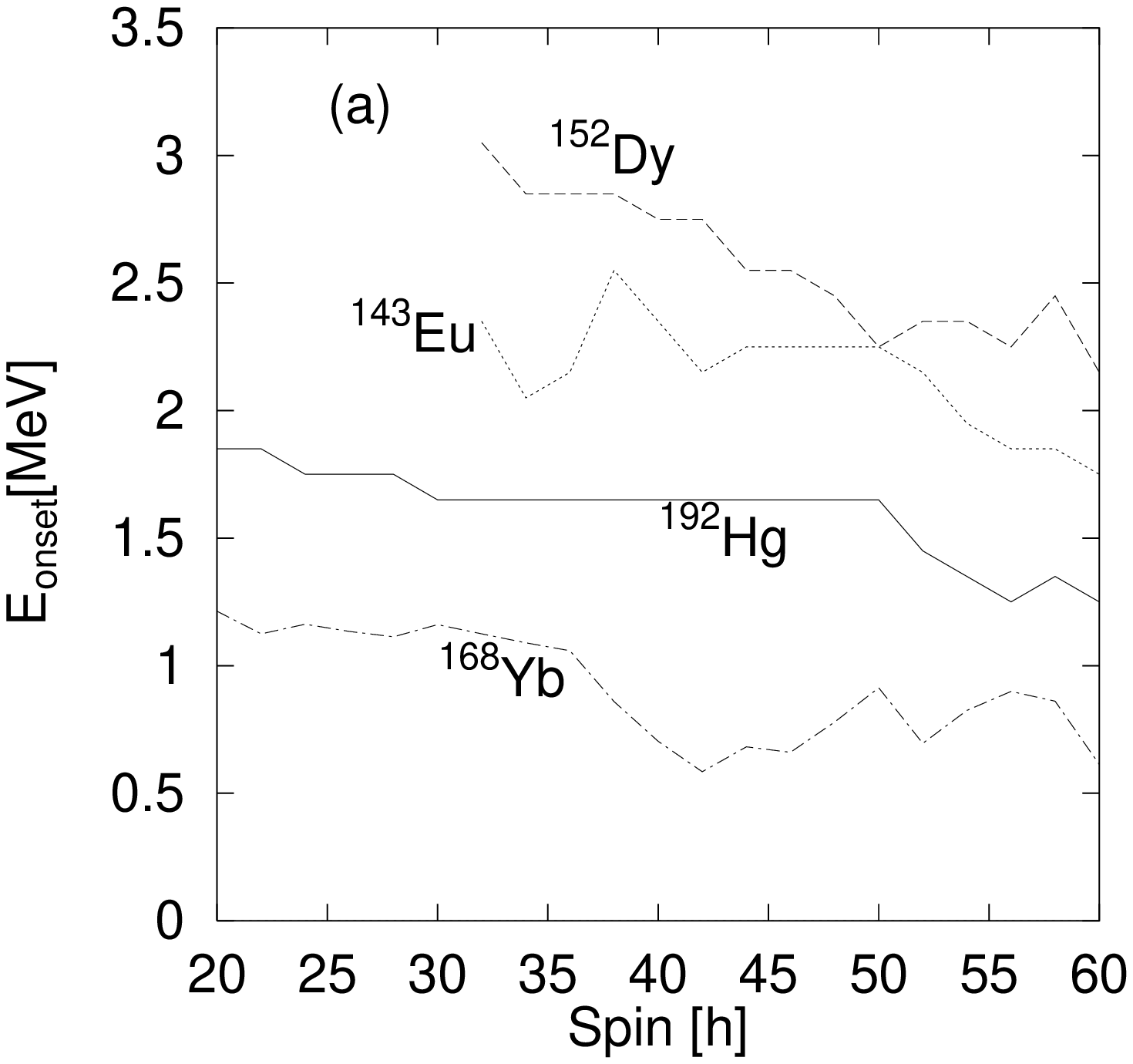}
\epsfxsize=70mm\epsffile{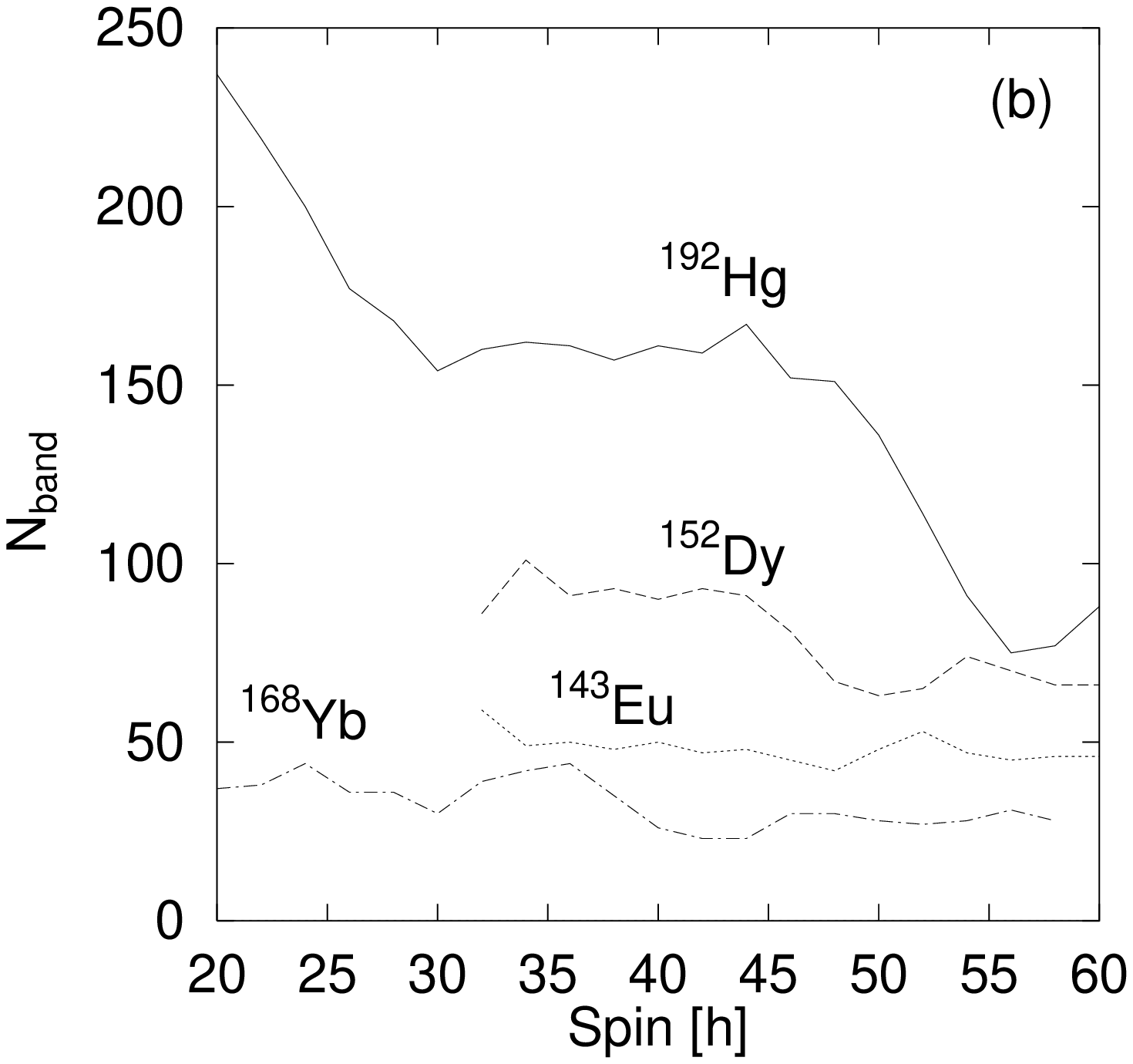}
}
\caption
{\label{number}
(a)The onset energy of rotational damping $E_{\rm onset}$ for
superdeformed 
$^{192}$Hg as a function of spin and  
those for superdeformed $^{152}$Dy, $^{143}$Eu and normally deformed
$^{168}$Yb. 
(b)The number of discrete
rotational bands $N_{\rm band}$ for the same nuclei.
}
\end{figure}

It is seen from Fig.\ref{flow} that there are apparently more than 20
rotational bands with even spin and parity in $^{192}$Hg, which lie
 between  the yrast line and up to $E_x\sim 1.5$ MeV above yrast. 
As the excitation energy increases, the rotational band structure
gradually 
disappears and the rotational damping sets in.
To describe this transition more quantitatively, we calculate an
effective number of E2 branches decaying from a state $\alpha$, which
is defined by 
\begin{equation}
n_{\rm branch}(\alpha)= \left( \sum_\beta w_{\alpha I\rightarrow
\beta I-2}^2 \right)^{-1}.
\end{equation}
Calculating the average value of $n_{\rm branch}$ as a function of
intrinsic excitation energy measured from the yrast line(by taking a
energy bin of 0.1 MeV interval), 
we define onset energy of damping $E_{\rm onset}$ by the condition that
$\langle n_{\rm branch}\rangle (E_{\rm onset}) \!=\!2$.
In Fig.\ref{number}(a), we show the obtained onset energy for $^{192}$Hg
as a function of angular momentum together with the results
for $A\sim 150$ superdeformed nuclei and normally deformed $^{168}$Yb. 
The onset energy varies in superdeformed $^{192}$Hg from 1.8 to 1.3
MeV with increasing 
spin. It is apparently smaller than those for $A\sim 150$
superdeformed nuclei 
and larger than $^{168}$Yb.

We define 
number of rotational bands denoted by $N_{\rm band}$ by counting at a
given spin the levels for which the condition $n_{\rm
branch}(\alpha)<2$ is satisfied for two consecutive decay-in
and decay-out E2 transitions. The numbers for four sets of signature
and parity quantum numbers are summed up.
This quantity corresponds to the experimental 
effective number of 
paths \cite{Herskind92} which
is extracted from the spectral fluctuation at the first ridge
of the $E_{\gamma_1}  \times  E_{\gamma_2}$ spectra, which are formed
by two consecutive
E2 transitions along rotational bands.
The calculated number of rotational bands 
is shown in Fig.\ref{number}(b) as a function of angular momentum.
A noticeable feature is that the number of superdeformed rotational
bands in $^{192}$Hg amounts up to about 150 on average, which is much
larger than in the $A\sim 150$ superdeformed nuclei (see also Fig.9 in
Ref.\cite{Yoshida97} ) and typical value 30 of $N_{\rm band}$ in
normally deformed nuclei.
Dependence of 
the number of bands
on spin is also distinct in $^{192}$Hg; it is about 250 at $I=20\hbar$
and decreases as spin 
increases and takes 
150 at $30\hbar \le I\le 46\hbar$ and again begins to decrease at
$I=48\hbar$ and reaches less than 100 at $I\ge 54\hbar$.

The quasi-continuum E2 $\gamma$ rays associated with superdeformed
states are observed in $^{192}$Hg in the spin region $I\sim
40\hbar$ \cite{Lauritsen92}, but there is no direct experimental
information, such as $N_{\rm band}$, concerning the onset of
rotational damping. 

\subsection{Rotational damping width}
\begin{figure}
\centerline{
\epsfxsize=58mm\epsffile{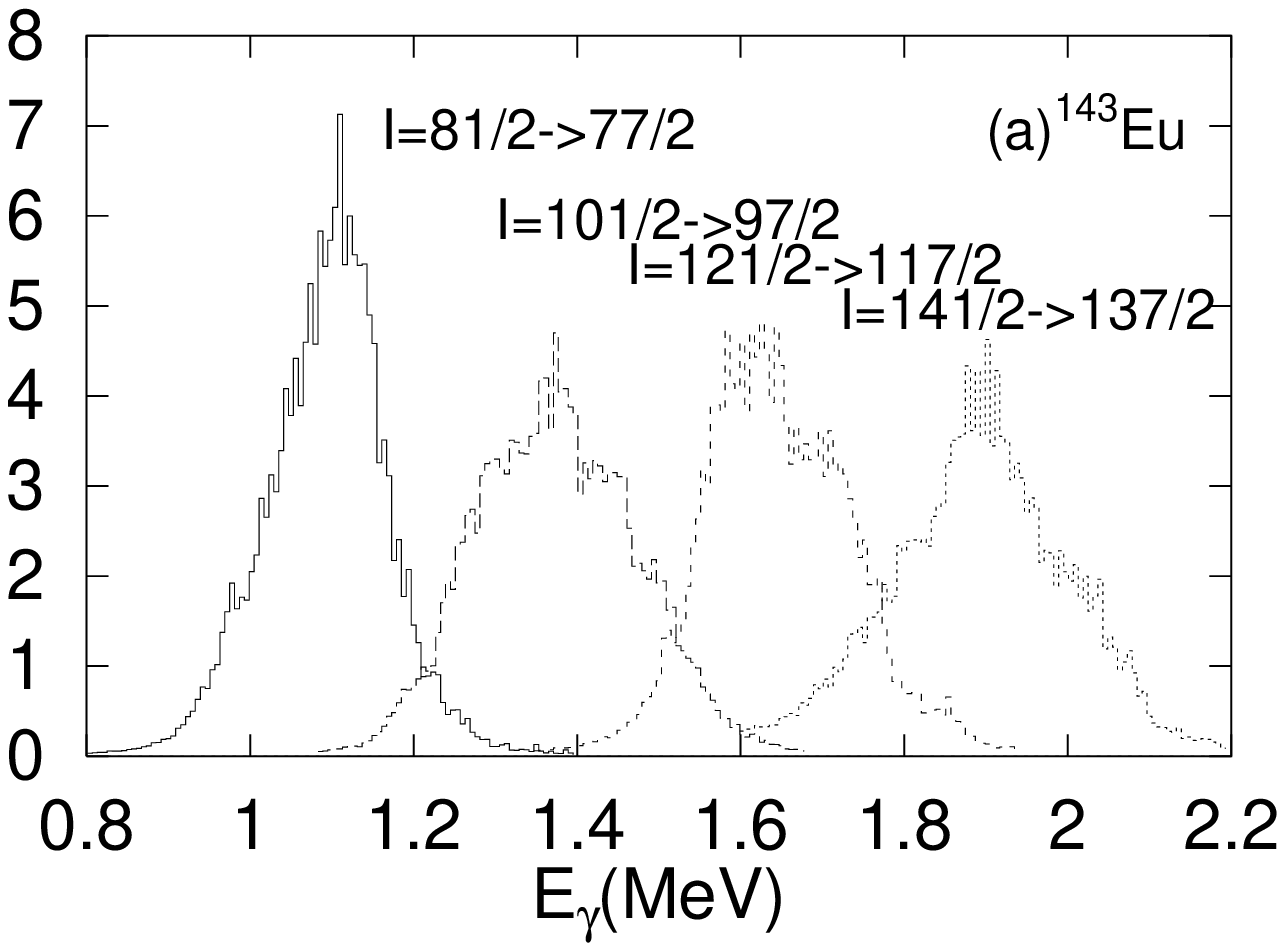}
\epsfxsize=54mm\epsffile{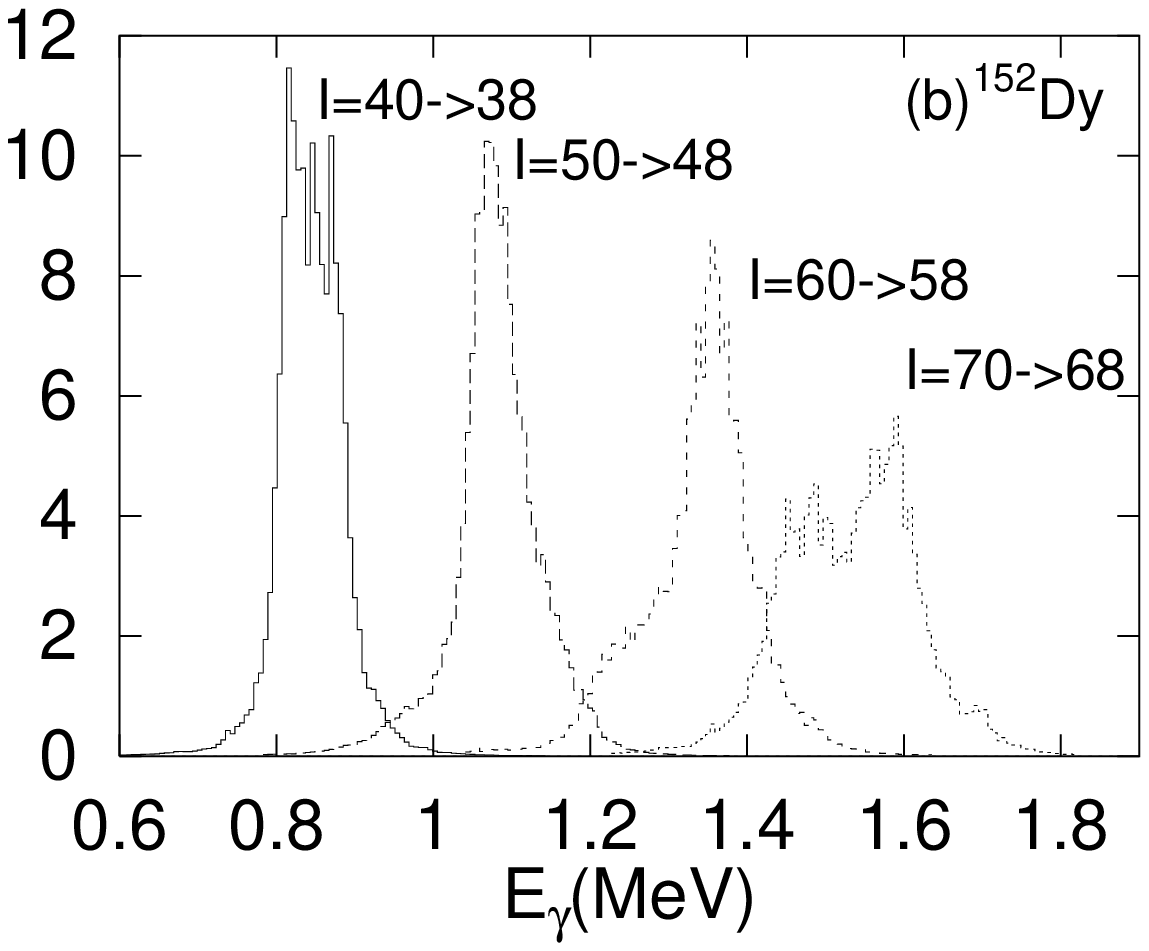}
\epsfxsize=44mm\epsffile{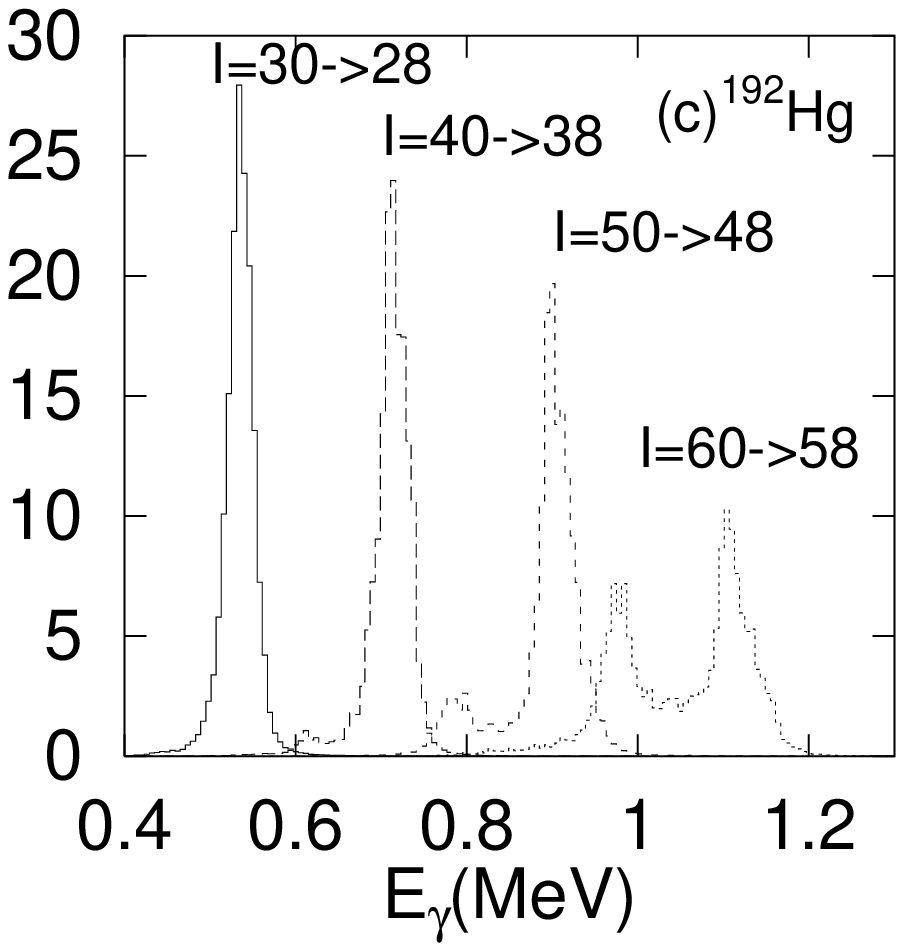}
}
\caption
{\label{strfn1}
The single-step E2 strength distribution $S^{(1)}(E_\gamma)$ for
transition 
$I\rightarrow I-2$ calculated for an energy bin composed of 
lowest 201-300th levels for each $I^\pi$ spectrum for superdeformed
$^{143}$Eu, $^{152}$Dy and $^{192}$Hg.
}
\end{figure}

In this subsection we discuss in detail
the damping width of collective rotational motion.
As excitation energy exceeds  the  onset energy of damping, E2 strength
decaying from an excited state is fragmented over several or more
final states and distributed within a certain interval of 
$\gamma$-ray energy $E_{\gamma}$, corresponding to the rotational
damping width  $\Gamma_{\rm rot}$.
The distribution is expressed by  a  single-step E2 strength function
defined by  
\begin{equation}
S^{(1)}(E_{\gamma})={\sum _{\alpha,\beta}}^{'}w_{\alpha I\rightarrow
\beta I-2}\delta(E_{\gamma}-E_{\alpha I}+E_{\beta I-2})/{\sum_\alpha}^{'} 
\end{equation}
where the summation of $\alpha$ runs over the initial states at $I$ in 
an energy bin, whereas $\beta$ runs over all final states at $I-2$.
Figure \ref{strfn1} shows the calculated E2 strength distribution
$S^{(1)}(E_\gamma)$
associated with an ensemble in an energy bin 
composed of 201st to 300th levels at each $I^\pi$ and averaged over
parities.
This ensemble  of levels
corresponds to the energy region, $E_x\approx$ 3.8, 3.6 and 
2.4 MeV above yrast for $^{143}$Eu, $^{152}$Dy and $^{192}$Hg,
respectively, which is
much higher than the onset energy by about 1-2 MeV.
The distribution for lower energy bins({\it e.g.}, those of
1st-100th  and  101st-200th) is not very different from that shown in
Fig.\ref{strfn1}.

The calculated strength distribution has the centroid corresponding to
$2\hbar\omega_{\rm rot}=\frac{2I}{\cal J}$(${\cal J}$ being the moment 
of inertia) and  the distribution width is  broadened with
increasing spin. The distributions in $^{152}$Dy and $^{192}$Hg are
split into two peaks at very high spins. This  splitting  comes from 
high-$j$ intruder orbits coming down close to the Fermi
surface with increasing spin. 
Since the configurations occupying the high-$j$
orbit $\pi 7_1$ for $^{152}$Dy or $\pi 8_1$ for $^{192}$Hg,
have larger angular momentum alignment by sizable amount than the
other ones,
transitions associated with  configurations occupying or
excluding the high-$j$ orbits  form
peaks at different $\gamma$-ray energies.
\begin{figure}
\centerline{
\epsfxsize=70mm\epsffile{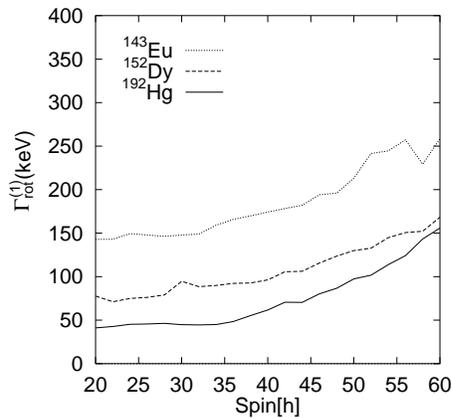}
}
\caption
{\label{width1}
The rotational damping width $\Gamma^{(1)}_{\rm rot}$ extracted from
standard deviation of the single-step E2 strength distribution
$S^{(1)}(E_\gamma)$ associated with the 201-300th levels, averaged over
both parities and signatures, for 
$^{192}$Hg and $A\sim 150$ superdeformed nuclei as  a function  of spin.
}
\end{figure}

It is useful to extract rotational damping width from the E2
distributions. However, it is difficult to define the FWHM because of
the two-peak structure of the distribution at higher spins. An
alternative definition may be given by 
the standard deviation $\sigma$ of the strength distribution
$S^{(1)}(E_\gamma)$ and we define
the width $\Gamma^{(1)}_{\rm rot}$ by $2\sigma$.
The evaluated $\Gamma^{(1)}_{\rm rot}$
averaged over both 
signatures and parities is shown in Fig.\ref{width1}. 
It is noted that $\Gamma^{(1)}_{\rm rot}$ in $^{192}$Hg is quite small 
$\Gamma_{\rm rot}^{(1)}\lesim 50$ keV at lower spins and increases with
increasing spin.
On the other hand 
$\Gamma^{(1)}_{\rm rot}$ for $^{143}$Eu is $150\sim 250$ keV, which is 
as large as for $A\sim 170$ normally deformed nuclei \cite{Matsuo97}.
That for $^{152}$Dy is $90\sim 150$ keV, which is sizably smaller than 
$^{143}$Eu though it is twice as large as that for $^{192}$Hg.
We remark that the
quantity $\Gamma_{\rm rot}^{(1)}$ may overestimate the rotational
damping width when the rotational damping is not well developed. For
example, if it is calculated for energy bins of near-yrast levels 
below the onset energy of damping, $\Gamma_{\rm rot}^{(1)}$
represents statistical dispersion of undamped E2 $\gamma$-ray energies 
which differ from band to band while the rotational damping width
should be zero.

Another way to extract the rotational damping width may be derived from
correlation between two consecutive rotational E2
transitions since the correlation distinguishes the damped transitions
and rotational band structure more clearly. This idea is adopted also
in the analysis of the experimental 
$E_{\gamma}\times E_\gamma$ spectra \cite{Rot-Map,Rev}.
The correlation  can be  represented by  a   function,
\begin{equation}
S^{(2)}(E_{\gamma 1}-E_{\gamma 2})={\sum_{\alpha\alpha '\alpha ''}}^{'}
w_{\alpha I\rightarrow \alpha ' I-2}w_{\alpha ' I-2\rightarrow \alpha
'' I-4}\delta(E_{\gamma 1}-E_{\gamma 2}-E_{\alpha I}+2E_{\alpha 
'I-2}-E_{\alpha ''I-4})/{\sum_\alpha}^{'}
\end{equation}
which we call the two-step strength function.
Here the summation runs over two consecutive transitions, {\it
i.e.}, the decay-in transitions from
$\alpha$ at $I$ to $\alpha^{'}$ at
$I-2$ and decay-out transitions from $\alpha^{'}$ 
to $\alpha^{''}$  at  $I-4$. 
 An  average is taken over levels $\alpha^{'}$ in an energy bin.
If we assume a simple Lorentzian distribution for the first transition
$I\rightarrow I-2$ and the second one $I-2\rightarrow I-4$,
$S^{(2)}$ 
become a convolution of the two Lorentzian distributions and the FWHM
of $S^{(2)}$ 
is equal to twice of the rotational damping width.
Thus it will be possible to define the rotational damping width as a
half value of FWHM of $S^{(2)}$. We use different notation
$\Gamma_{\rm rot}^{(2)}$ for this quantity 
to distinguish it from
$\Gamma_{\rm rot}^{(1)}$.

\begin{figure}
\centerline{
\epsfxsize=60mm\epsffile{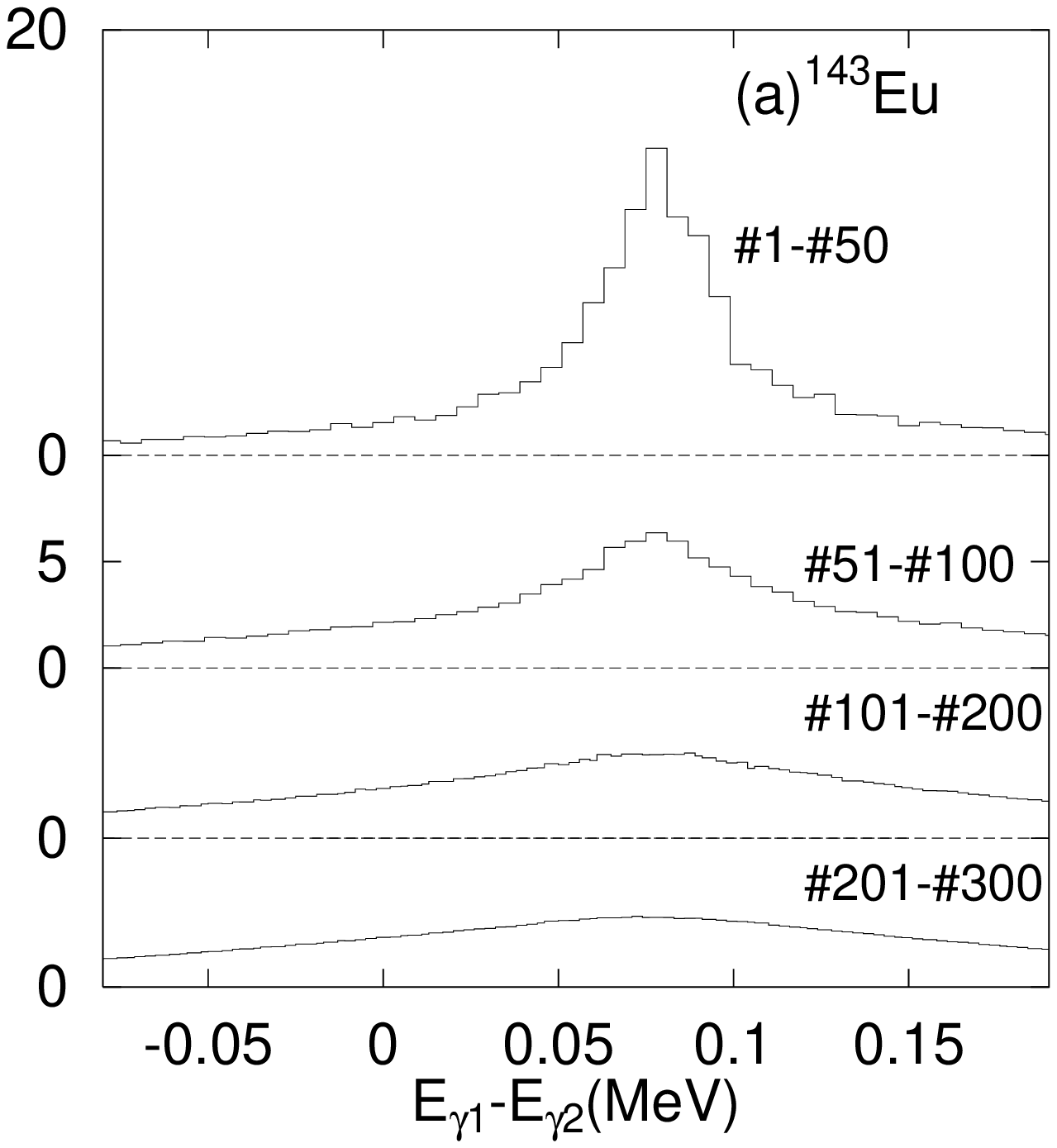}
\epsfxsize=60mm\epsffile{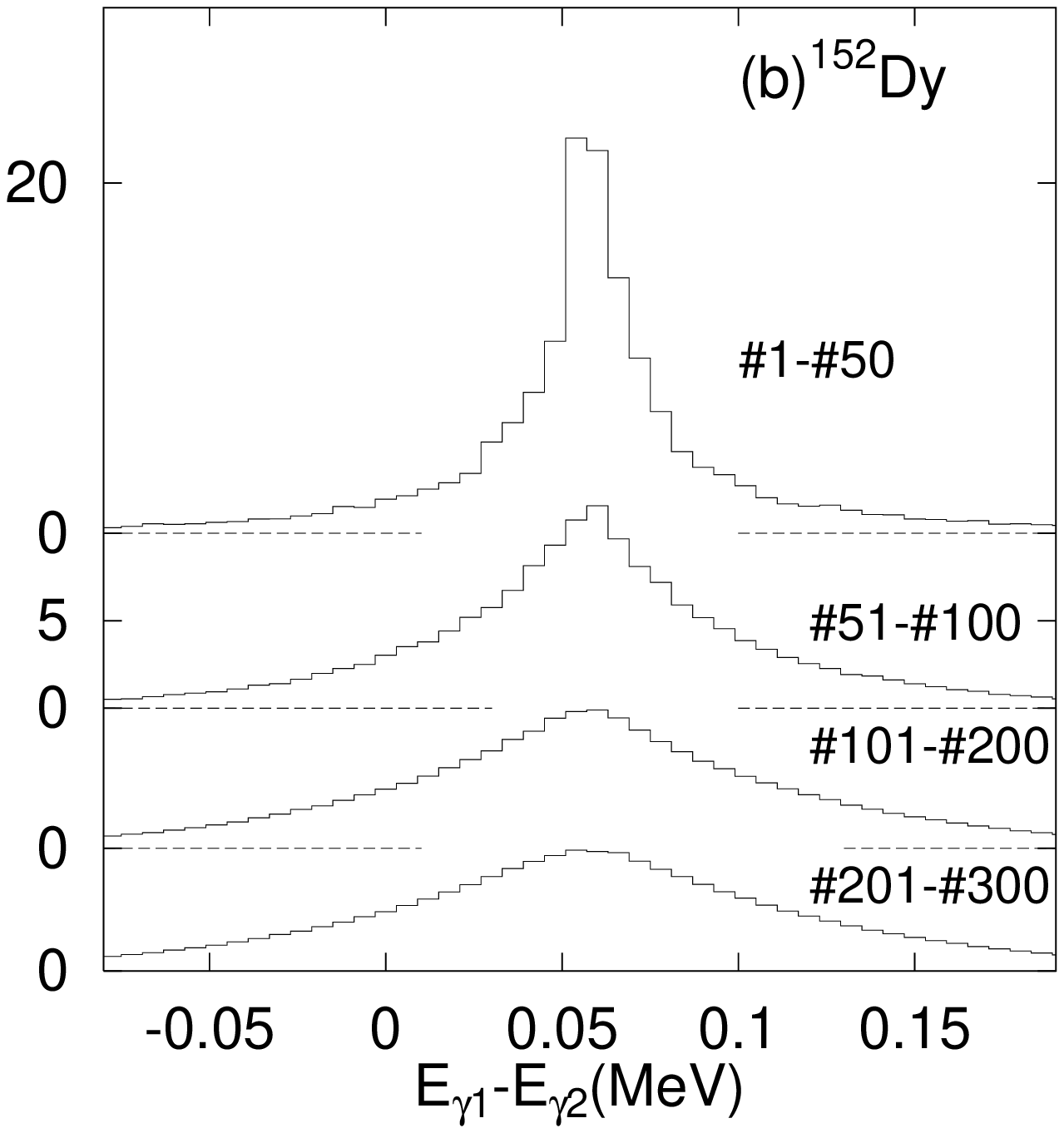}
\epsfxsize=60mm\epsffile{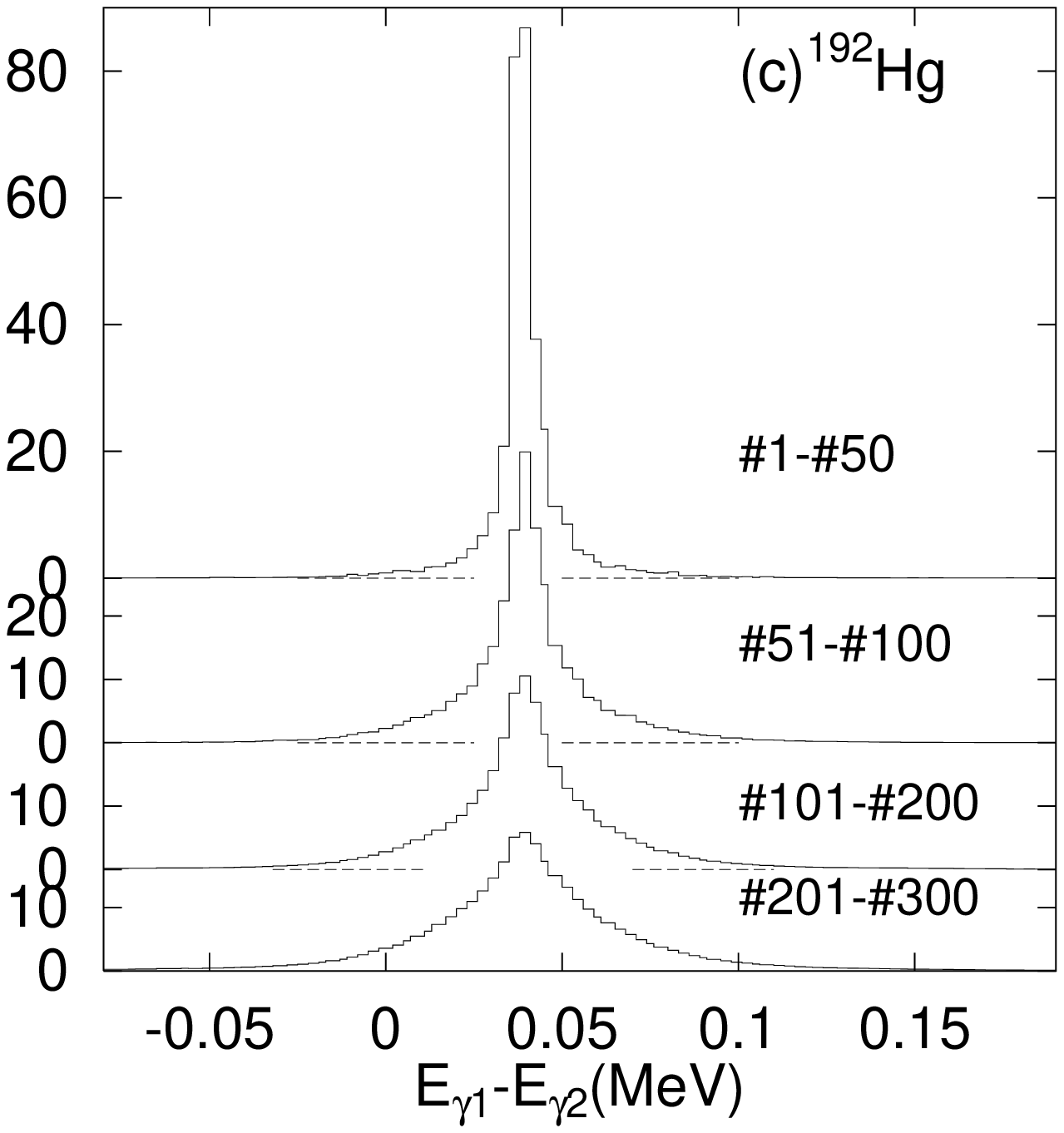}
}
\caption
{\label{strfn2}
The two-step E2 strength distribution $S^{(2)}(E_{\gamma
1}-E_{\gamma 2})$ associated with consecutive transitions 
$I\rightarrow I-2 \rightarrow I-4$ for different ensembles
composed of different levels. It is averaged over 
spin interval $I=34-47\hbar$ for  $^{192}$Hg, and
$I=44(89/2)-57(115/2)\hbar$ for $^{152}$Dy($^{143}$Eu). 
}
\end{figure}

\begin{figure}
\centerline{
\epsfxsize=70mm\epsffile{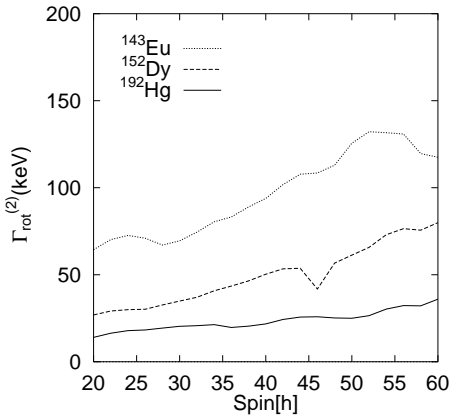}
}
\caption
{\label{width2}
The rotational damping width $\Gamma^{(2)}_{\rm rot}$ extracted from
FWHM of the two-step E2
strength distribution $S^{(2)}(E_{\gamma 
1}-E_{\gamma 2})$ associated with 201-300th levels as a function of spin.
}
\end{figure}

The calculated two-step strength function
$S^{(2)}$
shown in Fig.\ref{strfn2} does not necessarily show a simple Lorentzian 
nor Gaussian shape for the energy bin covering the onset region of 
damping. This is because there coexist undamped transitions which form 
rotational band  structure,  scars of 
rotational bands embedded in the damped region
(they contribute to the narrow
distribution) and fragmented damped transitions (producing wide
distribution with width of $2\times \Gamma_{\rm rot}^{(2)}$)
\cite{Leoni,Scar,Matsuo97}. As the excitation energy increases
the wide component dominates as seen for the
energy bin covering the 201st-300th levels. From this  energy bin  we
extract 
the rotational damping width $\Gamma_{\rm rot}^{(2)}$, which is
plotted in Fig.\ref{width2}. It is noted that the  two-step 
strength function $S^{(2)}$  
 has smooth  profile  in contrast to  the two-peak
behavior in the single-step strength function 
$S^{(1)}$. For this reason and the possible overestimate in
$\Gamma_{\rm rot}^{(1)}$, we refer mainly to $\Gamma_{\rm rot}^{(2)}$
as the rotational damping width $\Gamma_{\rm rot}$ in the following
discussion. 

It is seen from Fig.\ref{width2} that
the rotational damping width $\Gamma_{\rm rot}^{(2)}$
is $25\sim 30$ keV for $^{192}$Hg, $30\sim 70$ keV for $^{152}$Dy and
$70\sim 140$ KeV for $^{143}$Eu.  
Noticeable features are that  the  rotational damping width in $^{192}$Hg
is smaller by more than factor 4 than that in $^{143}$Eu, and that 
there is factor 2 difference between $^{143}$Eu and $^{152}$Dy in
spite of similar  mass number and deformation. These features
are also seen in $\Gamma_{\rm rot}^{(1)}$. So far there seems no
experimental analysis that extracts directly rotational damping width
associated with superdeformed states.

\section{Discussion}
\label{sec:discuss}
In this section we discuss in detail  the numerical results  shown
in the 
previous section with focuses on two distinctive
features found in superdeformed $^{192}$Hg, {\it i.e.},
(i) very small rotational damping width, $\Gamma_{\rm rot} \sim 20-30$
keV and 
(ii) quite large number of rotational bands, $N_{\rm band}\sim 150$.

\subsection{Small rotational damping width and shell effect
in alignments}
In order to understand the small rotational damping width in
$^{192}$Hg, it is 
instructive to recall the discussion by Lauritzen et.al
\cite{Lauritzen86} on the
basic features of rotational damping width. The unperturbed
basis rotational bands, {\it i.e.}, $n$p-$n$h configurations in the
cranked mean-field potential, have different rotational frequencies
at a given spin value,
depending on intrinsic particle and hole configurations. 
Once the residual interaction is taken into account, it causes 
mixing among the unperturbed rotational bands and an energy eigenstate 
becomes a superposition of different basis states at the same
spin. Under these circumstances, the rotational damping width is
estimated to be 
\begin{equation}
\label{ewidth}
\Gamma_{\rm rot}  \sim 4 \Delta \omega
\end{equation}
in terms of the statistical dispersion of E2 transition energies 
$\delta E_\mu=E_\mu(I)-E_\mu(I-2) \approx 2\omega_\mu$ associated with
basis bands $\mu$ or dispersion $\Delta \omega $
of rotational frequencies $\omega_\mu$ if the spreading width of basis
states (due to the 
configuration mixing) $\Gamma_\mu$ is relatively small
compared with the dispersion of rotational frequencies
$\Delta \omega$ ($\Gamma_\mu < 2 \Delta \omega$).
In the other case ($\Gamma_\mu > 2 \Delta \omega$), the rotational damping
width is estimated as 
\begin{equation}
\Gamma_{\rm rot}  \sim 2 \frac{(2 \Delta\omega)^2}{\Gamma_\mu}, 
\end{equation}
which is smaller than the estimate Eq.(\ref{ewidth}) ({\it i.e.}, the
motional 
narrowing). 
In any case, the damping width depends strongly (linearly or
quadruticly) on the dispersion of rotational frequency $\Delta \omega$.
 
\begin{figure}
\centerline{
\epsfxsize=70mm\epsffile{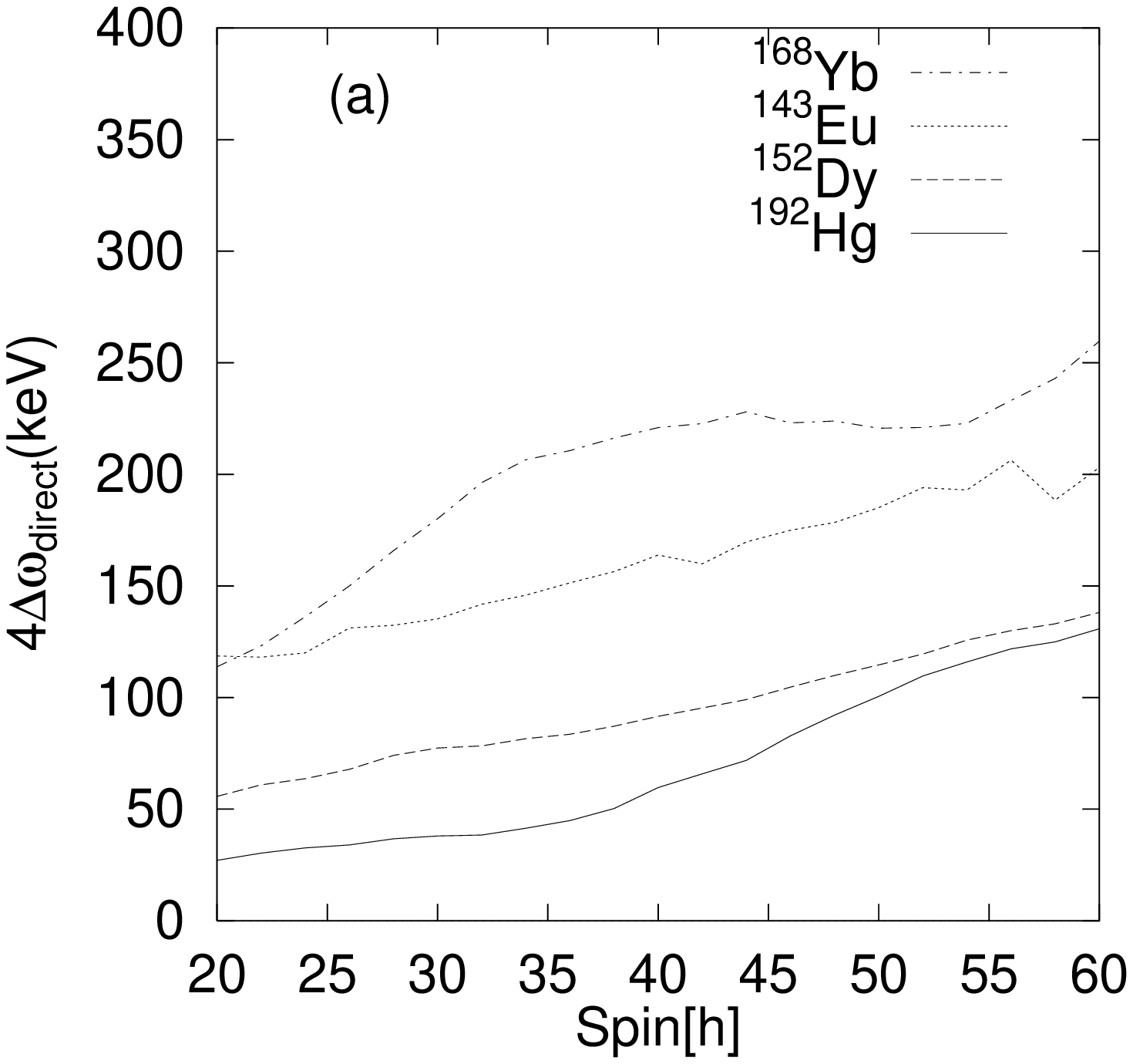}
\epsfxsize=70mm\epsffile{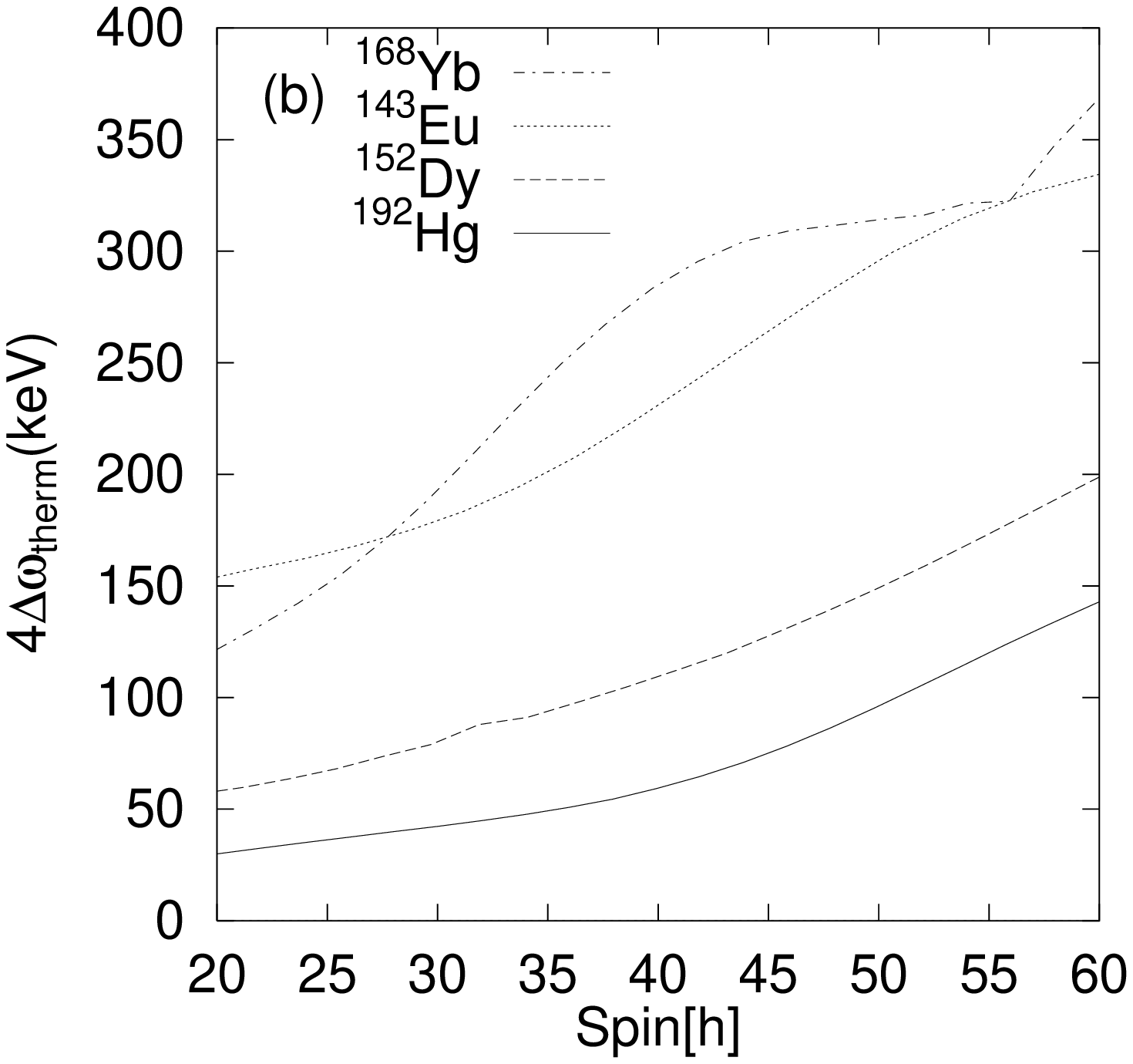}
}
\caption
{\label{disp-spin}
(a)The statistical dispersion of rotational frequency
$\Delta\omega_{\rm direct}$ which is calculated from the transition
energy $\delta E_\mu=E_\mu(I)-E_\mu(I-2)$ for the $n$p-$n$h states
covering the lowest 200 configurations for each $I^\pi$. Here
$4\Delta\omega_{\rm direct}$ is plotted as a function of spin.
(b)The dispersion of rotational frequency  $\Delta\omega_{\rm
therm}$  
obtained by means of the
thermal grand canonical
ensemble (Eq.(\protect \ref{thermal})). The temperature of the ensemble 
is chosen so that its average excitation energy agrees with that of
energy bin 
used to calculate $\Delta\omega_{\rm direct}$.
Namely,
$T=$0.31, 0.46, 0.48 and 0.32 MeV for $^{168}$Yb,
$^{143}$Eu, $^{152}$Dy and $^{192}$Hg, respectively.
}
\end{figure}

The dispersion of rotational frequency may be  evaluated   in two ways.
A direct way is just to calculate numerically the variance of
E2 transition energies $\delta E_\mu$ associated with
the basis states $\mu$ obtained in the microscopic calculation.
Namely, 
\begin{equation}
\label{direct}
\Delta \omega_{\rm direct} =  \frac{1}{2}\sqrt{\langle \delta E_\mu^2
\rangle - 
\langle\delta E_\mu\rangle^2} 
\end{equation}
where the average is taken over an ensemble of states in an energy bin.
In Fig.\ref{disp-spin}(a) we show the calculated dispersion(times
factor four) 
$4 \Delta \omega_{\rm direct}$
for the lowest 200 levels at each spin and parity
as a function of 
spin for the three superdeformed nuclei and a normally deformed nucleus
$^{168}$Yb. We plot  $4 \Delta \omega_{\rm direct}$ since it
corresponds to 
the simple analytic estimate of the rotational damping width 
(see Eq.(\ref{ewidth})).
The bin of lowest 200 levels has average excitation energy of 
$E_x=1.7, 2.7, 2.9$, and 1.8 MeV above the yrast line for  
$^{168}$Yb,
$^{143}$Eu, $^{152}$Dy, 
and $^{192}$Hg,  respectively.  It  covers the
onset region of rotational damping.

Another approximate  evaluation  can be made by introducing a
thermal grand canonical ensemble of excited configurations
in terms of cranked Nilsson orbits 
\cite{Nishinomiya}. This gives
\begin{equation}
\Delta\omega_{\rm therm}=\frac{1}{{\cal J}}\sqrt{\sum_n i_n^2f_n(1-f_n)},
\label{thermal}
\end{equation}
where it is assumed that nuclear shape is stable and the moment of
inertia  ${\cal J}$ does not depend very much on intrinsic
configurations (this can be justified in stably deformed nuclei).
In this case dispersion in the rotational frequency $\omega_\mu$ of
the $n$p-$n$h states is 
determined by the thermal fluctuation
in the alignment $\sqrt{\sum_n i_n^2f_n(1-f_n)}$ which is expressed in
terms of the 
single-particle alignments $i_n$ of the cranked Nilsson routhian
orbits and 
the Fermi-Dirac distribution function 
$f_n=(1+\exp{\frac{e_n-\lambda}{T}})^{-1}$ with $e_n$ and $\lambda$
being  the single particle routhian energy and the
Fermi energy.
Calculated dispersion $4\Delta\omega_{\rm therm}$ is shown in
Fig.\ref{disp-spin}(b). 
Here the temperature  $T$ is chosen so that average excitation energy 
$\langle E_x \rangle_{\rm therm}$ agrees with
the average energy of 
the ensemble shown in Fig.\ref{disp-spin}(a). 

Characteristic features of $\Delta\omega_{\rm direct}$ seen in 
Fig.\ref{disp-spin}(a) are: 1) $\Delta\omega_{\rm direct}$ varies quite
significantly  for   different nuclei.
It becomes the smallest
($< 50$ keV) in $^{192}$Hg at not very large spins ($I \lesim
40\hbar$), which is about four times smaller than that in
$^{143}$Eu and normally deformed  $^{168}$Yb. It
differs by about 
factor two between  superdeformed  $^{143}$Eu and $^{152}$Dy
although the 
mass number and 
deformation is rather similar. 2) $\Delta\omega_{\rm direct}$
increases as a function 
of spin (as can be expected from an analytic estimate \cite{Lauritzen86},
see also Eq.(\ref{harmonic})), but it exhibits strong non-linear
dependence on spin. 
It is noticed that $4 \Delta\omega_{\rm direct}$ shown in 
Fig.\ref{disp-spin}(a) agrees quite well with 
the ``rotational damping width''
$\Gamma^{(1)}_{\rm rot}$, shown in Fig.\ref{width1}, which is estimated
from the dispersion of the 
single-step E2 strength function $S^{(1)}(E_\gamma)$. 
The rotational damping width $\Gamma^{(2)}_{\rm rot}$
extracted from the two-step correlation (Fig.\ref{width2})
also reflects the above characteristic features.
The approximate estimate $\Delta\omega_{\rm therm}$ based on the grand
canonical ensemble, shown in 
Fig.\ref{disp-spin}(b), gives essentially the same results  as 
 $\Delta\omega_{\rm direct}$, 
and the two characteristic features discussed above are also seen. 
These observations indicate that the characteristic
behaviors of the dispersion $\Delta\omega$ of
rotational frequency and the rotational damping width $\Gamma_{\rm rot}$
can be ascribed to properties of the single-particle orbits of the
cranked Nilsson mean field, {\it i.e.}, the routhian
spectrum {$e_n$} and single-particle alignments {$i_n$}. 

\begin{figure}
\centerline{
\epsfxsize=70mm\epsffile{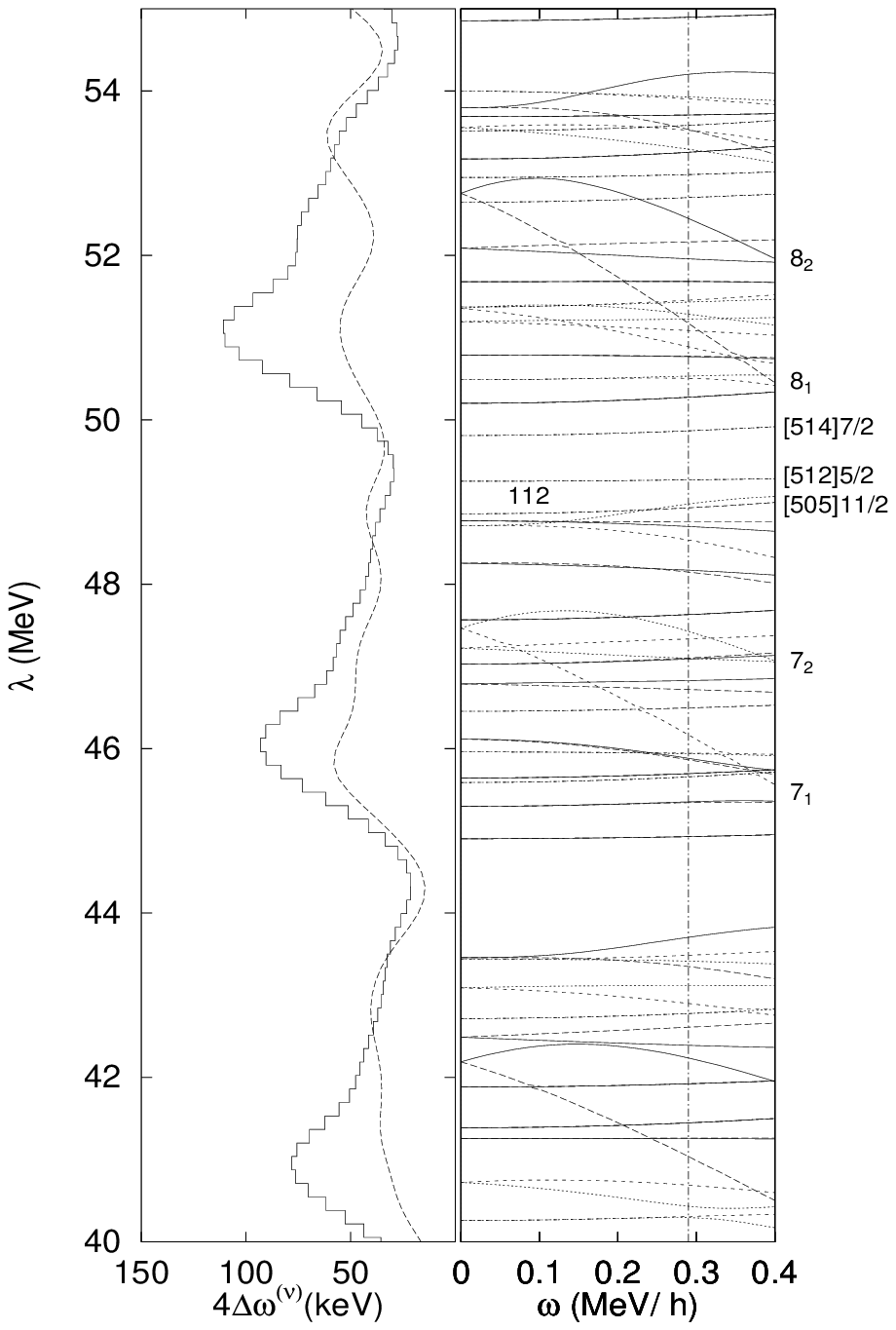}
\epsfxsize=70mm\epsffile{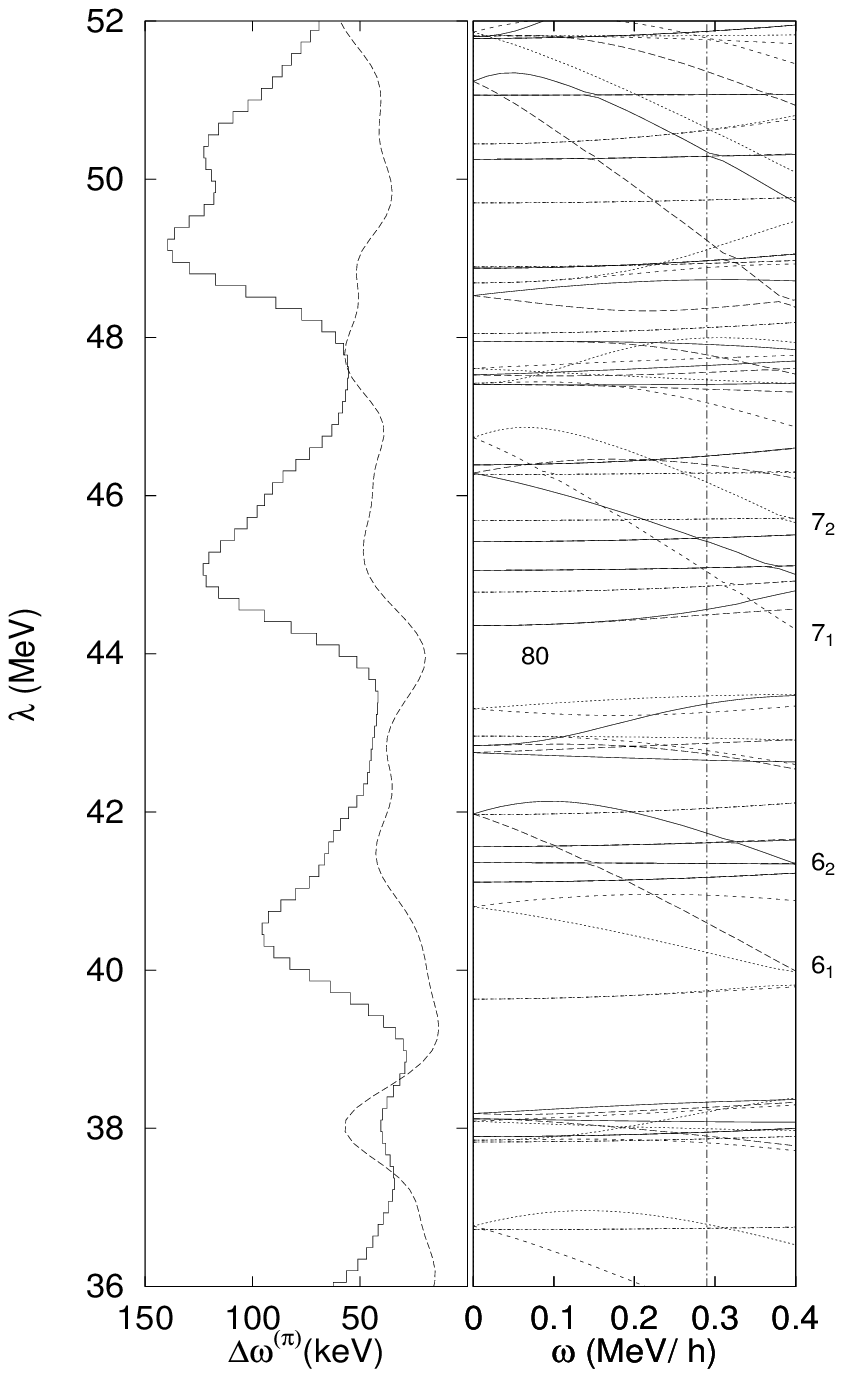}
}
\caption
{\label{smooth}
The neutron and proton contributions $\Delta\omega^{(\nu,\pi)}$ to the
dispersion of rotational frequency as a function of the Fermi energy
$\lambda$ for superdeformed $^{192}$Hg(solid line). 
Temperature is chosen
as $T=0.32$ MeV and rotational frequency $\omega=0.29{\rm
MeV}/\hbar$ corresponding to $I=40\hbar$.  
The smoothed spectral density(dashed line) is also drawn in an
arbitrary unit. The right panel shows the corresponding cranked Nilsson 
routhian diagram. The vertical line indicates $\omega=0.29$ MeV$/\hbar$.
}
\end{figure}

To clarify the relation of $\Delta\omega$ to the cranked Nilsson
single particle orbits, 
it is useful to introduce a quantity which represents 
density of the single-particle alignment
as a function of routhian energy
\begin{equation}
\rho_{\rm al}^{(\nu,\pi)}(e)  = \sum_{n \in \nu,\pi}i_n^2 \delta (e-e_n)
\end{equation}
which is defined separately for neutrons($\nu$) and protons($\pi$).
Here it is noted that 
the function  $\frac{1}{T}f_n(1-f_n) = \frac{d}{d\lambda} f_n$ 
in the r.h.s
of Eq.(\ref{thermal}) is a
smooth 
function peaked at the Fermi energy   $\lambda$ 
with width of order of temperature $T$ (and normalized as  $\int d\lambda
\frac{d}{d\lambda}f_n =1$). This allows 
us to regard the function  $\frac{d}{d\lambda}f_n \equiv g(\lambda -
e_n)$  as a
smoothing function. Introducing 
a smoothed density of the single-particle alignment by
\begin{equation}
\tilde{\rho}_{\rm al}^{(\nu,\pi)} (\lambda) = \sum_{n\in \nu,\pi}i_n^2
g(\lambda -e_n) 
\end{equation}
as a function of the Fermi energy $\lambda$,  the dispersion
$\Delta\omega_{\rm therm}$ 
of rotational frequency can be expressed in terms of the smoothed
single-particle alignment,
\begin{eqnarray}
\Delta \omega_{\rm therm}& = &\sqrt{ (\Delta \omega^{(\nu)})^2 +
(\Delta \omega 
^{(\pi)})^2},  \\
\Delta \omega^{(\nu,\pi)}& =& \frac{1}{\cal
J}\sqrt{T\tilde{\rho}^{(\nu,\pi)}_{\rm al}(\lambda_{\nu,\pi}) },
\label{dis-al}
\end{eqnarray}
where $\Delta \omega^{(\nu)}$ and $\Delta \omega^{(\pi)}$
 represent the neutron and proton
contributions to $\Delta \omega_{\rm therm}$, respectively.
They are plotted in 
Fig.\ref{smooth} as functions of the Fermi energy $\lambda$.

The quantity  $\Delta \omega^{(\nu,\pi)}$ (and the alignment density
$\tilde{\rho}_a^{(\nu,\pi)}$)  
shows   oscillation as a function of the Fermi energy with a typical
period of 5 MeV.
Maximum points of the oscillation are related to aligned high-$j$
or high-$N$ orbits. Taking $\Delta \omega^{(\nu)}$ as an
example, the 
maximum at $\lambda\sim 46$ MeV is caused by the lowest and most
aligned  
$N=7$ orbit ($7_1$) and the second $7_2$ orbit (which originate from 
$j_{15/2}$), and the next peak at $\lambda \sim 51$ MeV corresponds to
$8_1$ (and $8_2$). On the other hand, $\Delta \omega^{(\nu)}$ takes
locally 
minimum value at around $\lambda \sim 49$ MeV, where there exist
no highly aligned orbits, but instead present are
the oblate orbits 
with high-$\Omega$ value such as
$[514]\frac{7}{2},[512]\frac{5}{2},[505]\frac{11}{2}$  
which have little alignments ($ i_n \approx 0$). It should be noticed
that the neutron
Fermi 
surface of 
superdeformed  $^{192}$Hg($N=112$) is located near the minimum point of
the shell 
oscillation. The proton contribution $\Delta \omega^{(\pi)}$ also 
shows a similar shell oscillation, and the  proton Fermi surface of 
superdeformed $^{192}$Hg($Z=80$) is also situated near the minimum
position of 
the shell oscillation. 
In Fig.\ref{smooth}, the smoothed spectral density of single-particle
routhians $e_n$, {\it i.e.}, $\tilde{\rho}^{(\nu,\pi)}(\lambda) =
\sum_{n\in \nu,\pi}
g(\lambda -e_n)$, 
is also plotted. From the comparison with
$\tilde{\rho}_{\rm al}^{(\nu,\pi)}$, it is seen 
that the shell oscillation of the alignment density is not
directly related to the spectral density, but mainly reflects 
the single-particle alignments {$i_n$}. 

One can easily expect that 
the maxima and minima of the alignment density 
(dispersion of rotational frequency) varies  strongly as a function
not only of the Fermi energy but also 
of deformation and rotational frequency
since the routhian energy of the strongly aligned orbits generally
depends on these quantities. 
The non-linear spin dependence of $\Delta\omega$ seen in 
Fig.\ref{disp-spin} is originated from the shell effect
in the single-particle alignments. 
 Similar non-linear spin dependence 
due to aligned orbits was also recognized in the calculation of  
$\Delta\omega_{\rm therm}$ in normal deformed
$^{168}$Yb \cite{Nishinomiya,Herskind93}. 
\begin{figure}
\epsfxsize=70mm\epsffile{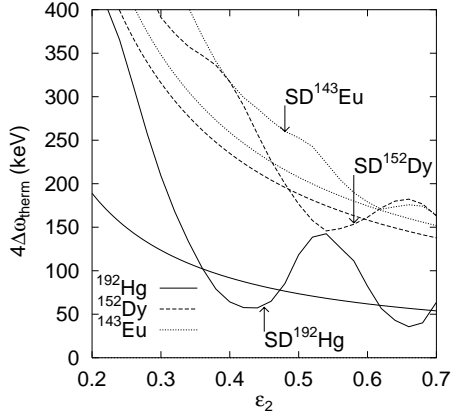}
\caption
{\label{disp-def}
The dispersion of rotational frequency $4\Delta\omega_{\rm therm}$
as a function of quadrupole
deformation 
parameter $\epsilon_2$  at   $I=40\hbar$ for $^{192}$Hg and $I=50\hbar$
for the others.
Temperature is the same as Fig.\protect\ref{disp-spin}.
The analytic estimate  $4\Delta\omega_{\rm h.o.}$ showing the
$\epsilon_2^{-1}$ dependence is also plotted for the sake of comparison. 
The equilibrium deformation of superdeformed states is indicated by
arrows. 
}
\end{figure}
 The dispersion
$\Delta \omega_{\rm therm}$ of rotational frequency calculated as a function
of quadrupole deformation parameter(Fig.\ref{disp-def})  also  shows
significant oscillation. 
In Fig.\ref{disp-def}, we  plot for comparison an analytic estimate
\cite{Lauritzen86,Nishinomiya} 
of $\Delta\omega$ based on the
harmonic potential model 
\begin{equation}\label{harmonic}
4\Delta\omega_{\rm h.o.}= 290.2  A^{-29/12}\epsilon^{-1}I E_{\rm
ex}^{1/4}[{\rm MeV}],
\end{equation}
which neglects the shell effects and therefore gives an average (and
smooth) 
behavior. 
The comparison indicates that the shell effect of alignment
is the same order of magnitude as the average value.
In $^{192}$Hg the shell effect suppresses the dispersion of rotational
frequency at the deformation (and Fermi surface)
of the superdeformed states. Note that different situation is realized in
superdeformed nuclei in $A 
\sim 150$ region, 
where the Fermi surface is located in the vicinity of  
the aligned high-$N$ orbits such as $\nu 7_1$ and $\nu 7_2$.
This enhances $\Delta\omega$ in $^{143}$Eu as shown in
Fig.\ref{disp-def}.

Thus the shell structure in the single-particle alignment influences
significantly the dispersion of rotational
frequency $\Delta\omega$, and consequently this causes very small
rotational damping 
width 
$\Gamma_{\rm rot}$ in superdeformed $^{192}$Hg.
\subsection {Large number of bands and retarded onset of damping}
In this subsection, we discuss the mechanism for
the very large number of rotational superdeformed bands($N_{\rm
band}\sim 150$) obtained in
$^{192}$Hg. 

In the preceding paper \cite{Yoshida97}, we showed 
that the relatively large number of bands $N_{\rm band}\sim 40-100$
seen in the calculation for $A\sim 150$ superdeformed nuclei(compared 
with the value $N_{\rm band}\sim 30$ for rare-earth normally deformed 
nuclei) is due to the
level density effect associated 
with the superdeformed shell gap in the single-particle routhian
spectrum. This mechanism, however, does not account for the
calculated results for $^{192}$Hg.

To show this, let us repeat briefly the arguments for the level density
effect \cite{Yoshida97}. In that argument we assumed 
that the onset of 
rotational damping takes place
 as soon as  the configuration mixing among basis
$n$p-$n$h  
states becomes strong. The condition for configuration mixing
may be expressed  as  $\Gamma_\mu > 1/\rho_{\rm 2body}$ in terms of 
the spreading width $\Gamma_\mu$ of the basis
$n$p-$n$h states  
and the density $\rho_{\rm 2body}$  of $n$p-$n$h states which can
interact via 
the two-body  interaction \cite{Lauritzen86}.  Assuming the
Fermi golden rule for the spreading width 
$\Gamma_\mu=2\pi\rho_{2\rm body}\bar{v}^2$, the
condition can be written also in terms of the average matrix element
$\bar{v}$ 
of the residual two-body interaction and the spacing 
 $d_2=1/\rho_{\rm 2body}$  
between
interactive  states,
\begin{equation}
\label{cond2}
d_2<\sqrt{2\pi}\bar{v}.
\end{equation}
If we assume the level density
$\rho$ and two-body level density $\rho_{\rm 2body}$ of $n$p-$n$h
states follow the Fermi
gas formula \cite{Bohr-Mottelson69I,Aberg88}, 
\begin{equation}
\label{fermtot}
\rho(E)=\frac{\sqrt{\pi}}{48}a^{-\frac{1}{4}}
E^{-\frac{5}{4}}\exp{2\sqrt{aE}},
\end{equation}
\begin{equation}
\label{twobody}
\rho_{\rm 2body}(E)=\frac{81}{4 \pi^6}a^{5/2}E^{3/2},
\end{equation}
the onset energy of configuration mixing (and rotational damping) is
expressed in terms of the level density parameter $a$ and the average
matrix element $\bar{v}$ as
$E_{\rm onset}\propto
\bar{v}^{-2/3}a^{-5/3}$ \cite{Yoshida97}. Similarly 
the number of rotational bands is expressed as 
$N_{\rm band} = \int^{E_{\rm onset}} \rho dE$,  
 which is a monotonically increasing function of  $aE_{\rm onset}$ 
and hence of $(a\bar{v})^{-1}$.  

\begin{figure}[htbp]
\centerline{
\epsfxsize=70mm\epsffile{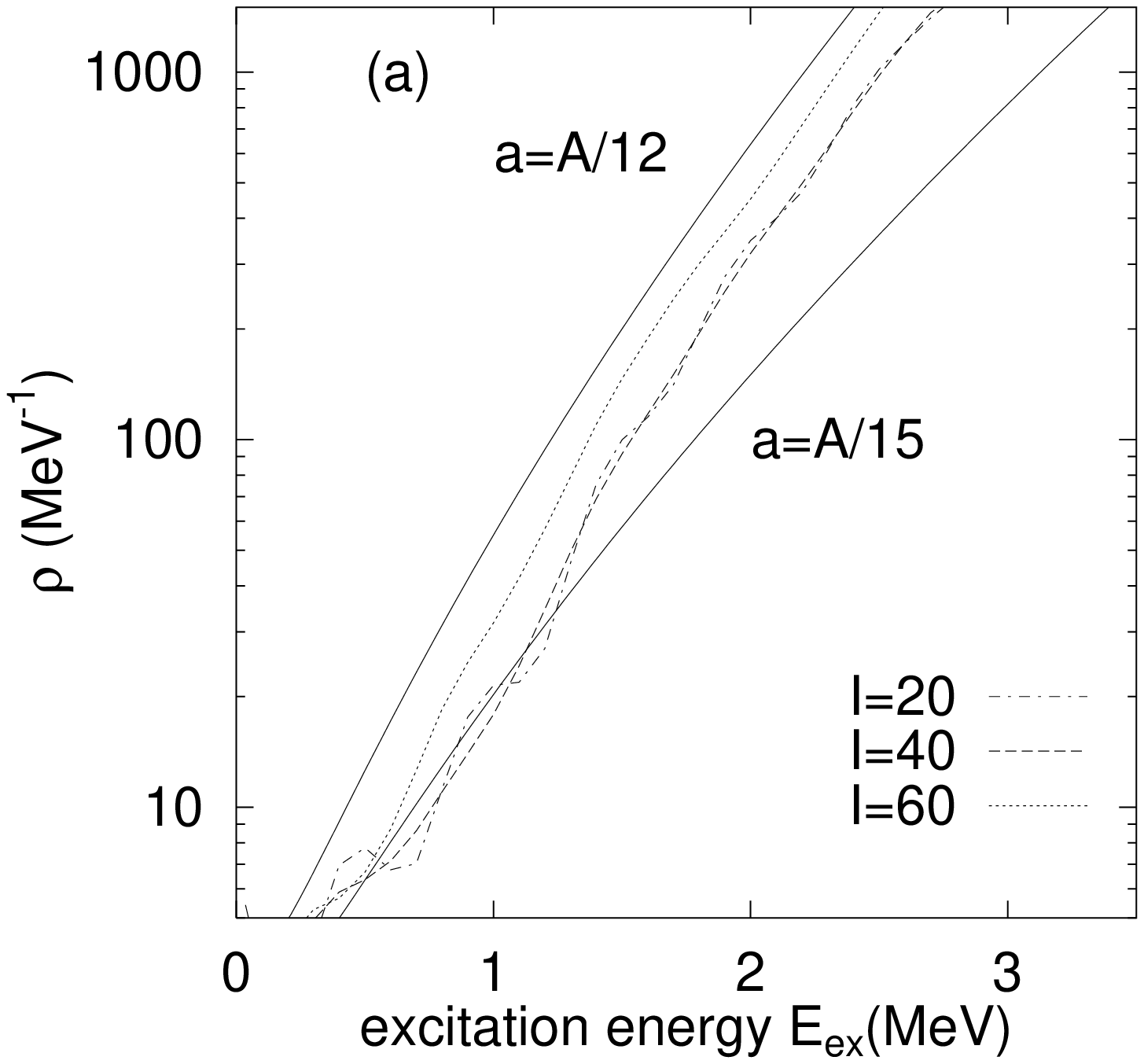}
\epsfxsize=70mm\epsffile{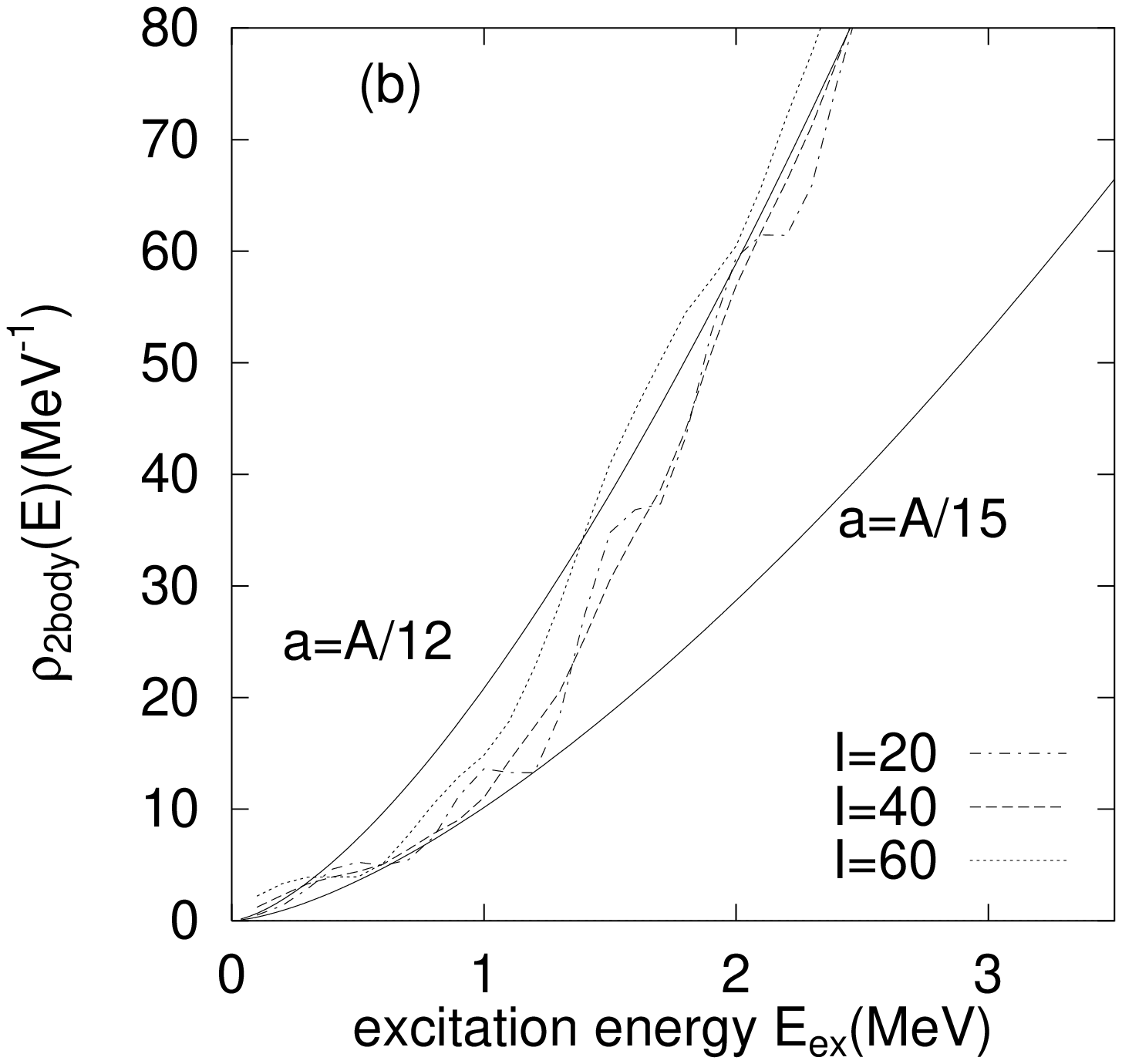}
}
\caption
{\label{density}
(a)The total level density and (b)the two-body level density of
unperturbed 
many-particle many-hole states  for superdeformed
$^{192}$Hg at spins $I=20,40,60\hbar$.
Here average is taken over both parities and signatures, {\it i.e.},
$I^{\pm},I+1^{\pm}$. The Fermi gas formulas with the level density
parameters $a=A/12$ and $A/15$ are also plotted.
}
\end{figure}
The level density parameter $a$ is usually parameterized  as $a=A/a_0$
where $a_0\sim 10$ MeV for normally deformed nuclei in the cranked
Nilsson model. For the superdeformed states, the level density
is smaller than that of normally deformed states
because of the shell gap in the single-particle routhian spectrum, and
this can be represented by larger value of $a_0$, which is estimated as
$a_0=15, 17$ MeV for $^{143}$Eu and
$^{152}$Dy, respectively \cite{Yoshida97}.  
Calculated level densities of superdeformed states for $^{192}$Hg is
shown in Fig.\ref{density}, from which 
$a_0 \sim 13$ MeV is estimated.
On the other hand, the average magnitude $\bar{v}$  of the two-body
elements $v_{ijkl} = \int d\vec{x}d\vec{x}'v_\tau\delta(\vec{x}-\vec{x}')
\phi_i^*\phi_j^*\phi_k\phi_l$
is estimated as  $\bar{v}
\sim$ (volume integral of the 
two-body force)/(nuclear volume) $\propto A^{-1}$. The r.m.s value of
the calculated off-diagonal 
matrix elements $v_{ijkl}$ is $\bar{v}=$20, 18, 12 and 16 keV for
$^{143}$Eu, 
$^{152}$Dy, $^{192}$Hg, and normally deformed  
$^{168}$Yb. They approximately follow the expected $A^{-1}$
dependence. 
Using the above characterizations of the level density and the
residual interaction, the dependence of $E_{\rm onset}$
and $N_{\rm band}$ on the level density shell effect (parameter $a_0$) 
and the mass number $A$ is estimated as 
$E_{\rm onset}\propto A^{-1}a_0^{-5/3}$, and
$N_{\rm band}$ is an monotonically increasing function of $a_0$
and independent of $A$.  This argument predicts that $N_{\rm band}$ in
$^{192}$Hg would be smaller than those in $^{152}$Dy and $^{143}$Eu.

\begin{figure}
\centerline{
\epsfxsize=70mm\epsffile{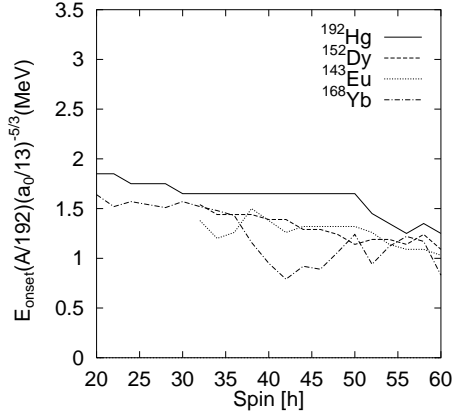}
}
\caption
{\label{reonset}
The onset energy of rotational damping $E_{\rm onset}$ for superdeformed
$^{192}$Hg,
$^{152}$Dy, $^{143}$Eu and normally deformed $^{168}$Yb multiplied by
a factor $(A/192)(a_0/13)^{-5/3}$, which removes the level density
effect. 
We adopt $a_0=13$ for $^{192}$Hg and
$a_0=17,15,10$ for  $^{152}$Dy, $^{143}$Eu and $^{168}$Yb,
respectively.
}
\end{figure}

It is noted, however, that 
microscopically calculated $N_{\rm band}$ for superdeformed
$^{192}$Hg is much larger 
than those for $^{152}$Dy and $^{143}$Eu (Fig.\ref{number}(a)).
This indicates  that the
excessively large number of superdeformed rotational 
bands in $^{192}$Hg cannot be explained just by the shell
effect in the level density. This is also demonstrated in
Fig.\ref{reonset} which 
plots the microscopically calculated $E_{\rm
onset}$  multiplied by a
factor $(A/192)(a_0/13)^{-5/3}$ with which we intend to remove 
the level density shell effect. It is seen that the 
scaled onset energy in $^{192}$Hg is  higher than that for $^{143}$Eu,
$^{152}$Dy and $^{168}$Yb. If we assume the above argument 
the onset energy of configuration mixing is estimated as
$E_x =
1.2\sim 1.3$ MeV for $^{192}$Hg with use of
the scaling property $\propto A^{-1} a_0^{5/3}$ and $E_{\rm 
onset} = 800$ keV and $a_0=10$ MeV
\cite{Matsuo97} for normally deformed $^{168}$Yb whereas the
microscopically 
calculated 
onset energy of rotational damping is about 1.6 MeV.
  
The above analysis indicates that the onset of
rotational damping  in $^{192}$Hg 
takes place at higher excitation energy ($E_x\sim 1.6$ MeV) than onset
energy ($E_x\sim 1.2$ MeV) of configuration mixing of $n$p-$n$h states.
Namely, the onset of
rotational damping is slightly "retarded" in comparison 
 with the onset of
 mixing. This  indicates that 
the mechanism of the onset of rotational damping is
qualitatively different from those in the normal
deformed rare-earth nuclei and $A \sim 150$ superdeformed nuclei.

To understand the mechanism of onset of rotational damping in
$^{192}$Hg, it is useful  
to recall the argument \cite{Mottelson93,Aberg96} that discusses 
possible occurrence of
'ergodic rotational bands' which means rotational
band structures formed by 
compound intrinsic states (the states whose wave functions have
strong configuration mixing). 
The ergodic
rotational bands are expected to realize when 
the rotational damping width is smaller than the level spacing
$d=1/\rho$ of the all levels at fixed $I^{\pi}$,
\begin{equation}
\label{ergodic2}
\Gamma_{\rm rot} < d,
\end{equation}
while wave functions of the energy levels are compound mixture
of basis many-particle many-hole states.
In this situation, a compound state at spin $I$ decays by E2
transition to a single
final compound state at $I-2$, 
and hence sequence of such E2 transitions forms a band structure in
spite of the strong mixing.

In the calculated situation for $^{192}$Hg, many of the excessive
number of rotational bands reside 
in the energy region between $E_x \sim 1.2$ MeV (the estimated onset
energy of mixing, see above) 
and $E_x \sim 1.6$ MeV (the onset energy of rotational damping).
Many of the energy levels in this energy region are some mixture
of basis many-particle many-hole configurations although
they are not fully ergodic since their excitation energy is
just a few hundred keV above the onset energy of configuration mixing
\cite{LevelStatistics}. 
The rotational damping width $\Gamma_{\rm rot}$ in $^{192}$ Hg is
evaluated to 
be 20-30 keV for
spins $I <50$(Fig.\ref{width2}).  The total level spacing $d$ is 
 $\sim 30$ keV 
at $E_x=1.2$ MeV,  and $\sim 10$ keV at $E_x=1.6$ MeV
(estimated from Fig.\ref{density}(a)). Namely,  the rotational damping 
width is comparable with the level spacing, 
\begin{equation}
\label{rot-den}
\Gamma_{\rm rot} \sim d \,
\end{equation}
in the energy region under discussion. Thus, the situation lies
around the
borderline of the condition Eq.(\ref{ergodic2}) for
the ergodic rotational bands.
This suggests that similar
mechanism to that of the ergodic 
rotational bands is taking place in superdeformed $^{192}$Hg, 
and explains why 
the energy levels in the region of $E_x \sim 1.2-1.6$ MeV
form
rotational band structures and contribute to
the large number 
of $N_{\rm band}$. The above argument 
 explains also the observation that $N_{\rm
band}$  in $^{192}$Hg decreases as spin increases,
especially for $I > 50$, since $\Gamma_{\rm rot}$ becomes as large as 
$\Gamma_{\rm rot} \gesim 30$ keV for $I\ge 50\hbar$, which tends to
violate the relation Eq.(\ref{rot-den}).
Note that the above mechanism does not manifest itself in
normally  deformed 
$^{168}$Yb and superdeformed $^{143}$Eu where $\Gamma_{\rm rot}$
is much larger. For superdeformed $^{152}$Dy, the mechanism may be 
relevant to explain the small spin dependence of $N_{\rm band}$ seen in
Fig.\ref{number}(b) although the 
effect is not very  
significant (cf Ref.\cite{Yoshida97}).
It should be recalled here that the shell effect associated with the
single-particle alignment 
plays an important role in causing such small $\Gamma_{\rm rot}$
particularly in superdeformed $^{192}$Hg, but not 
in other calculated nuclei, as we discussed in the previous
subsection. 

\section{Conclusions}
\label{sec:conclude}
We performed microscopic description of the rotational damping
associated with superdeformed states in $^{192}$Hg.
We obtained remarkably large number of superdeformed rotational bands,
$N_{\rm band}\sim 150$ and very small rotational damping width  
$\Gamma_{\rm rot}\sim 30$ keV, which are different by more than
factor 2 from those in 
$A\sim 150$ superdeformed nuclei and rare-earth normally deformed
nuclei. 
It is also found that the rotational damping width varies in different 
nuclei in the $A\sim 150$ region.
The strong variation of the rotational damping width is attributed to
the shell structure associated with the angular momentum alignment of
the single-particle orbits in the cranked Nilsson potential.
The alignments of the excited many-particle many-hole configurations
have different values for different states.
Their dispersion, which is strongly related to the rotational damping
width, depends on the presence of highly aligned
orbits in the vicinity of the Fermi surface.

We pointed out that in superdeformed $^{192}$Hg the very small
rotational damping width due to the shell effect causes slightly
different mechanism for the onset of rotational damping. In the cases
of the rare-earth normally deformed nuclei and $A\sim 150$ superdeformed 
nuclei, the onset of rotational damping takes place at the same energy
as the onset energy of configuration mixing among $n$p-$n$h 
states.  In $^{192}$Hg, on the other hand, energy levels in the
onset region 
of configuration mixing do not necessarily accompany damped
rotational decay since the damping width is as small as the
level spacing  
of the final states. This leads to the significantly large number of
superdeformed rotational bands in $^{192}$Hg.

Summarizing the present investigations together with those in the
preceding paper \cite{Yoshida97}, we revealed 
two kinds of shell effects which affect properties of the rotational
damping associated with superdeformed states.  Firstly,  because of the
shell gap in the single-particle spectrum of the rotating
superdeformed mean field, the level density of excited superdeformed
states is relatively low and the configuration mixing is weak compared 
with normally deformed nuclei. This influences onset properties of the
rotational damping as demonstrated by the calculations for $A\sim 150$ 
 region \cite{Yoshida97}. 
The second one, which was discussed in the present paper, is
originated from the shell structure in the angular momentum alignments
of the single-particle orbits. It affects strongly the damping
width  of rotational damping, but also modifies the onset properties of
the damping under 
certain conditions as seen in $^{192}$Hg.

\vspace{10mm}
\section*{Acknowledgment}
The authors 
 would like to thank  K. Matsuyanagi, T. D\o ssing, and E. Vigezzi
 for
helpful discussions.

\end{document}